# Yield scalings of clusters with fewer than 100 nucleons

James B. Elliott*, Kyrill A. Bugaev*, Luciano G. Moretto* and Larry Phair*

*Nuclear Science Division, Lawrence Berkeley National Laboratory, Berkeley, California 94720

**Abstract.** This document gives a historical review of the scaling of particles yields emitted from excited nuclei. The focus will be on what scaling is, what can be learned from scaling, the underlying theory of why one might expect particle yields to scale, how experimental particle yields have been observed to scale, model systems where particle (cluster) yields do scale and finally scaling observed in the particle yields of various low and medium energy nuclear reaction experiments. The document begins with a basic introduction to scaling in the study of critical phenomena and then reviews Fisher's theory which has all the aspects of scaling and can be directly applied to the counting of clusters, the most reliable measurement accessible to the experimental study of nuclear reactions. Also this document gives a history of the various scalings observed in nuclear reaction experiments and culminates with an estimate of the nuclear liquid-vapor phase boundary based upon measured particle yields.

## INTRODUCTION

This document performs the modest task of covering a century worth of research on scaling in condensed matter and nuclear physics [1, 2, 3, 4, 5, 6, 7, 8, 9, 10, 11, 12, 13, 14, 15, 16, 17, 18, 19, 20, 21, 22, 23, 24, 25, 26, 27, 28, 29, 30, 31, 32, 33, 34, 35, 36, 37, 38, 39, 40, 41, 42, 43, 44, 45, 46, 47, 48, 49, 50, 51, 52, 53, 54, 55, 56, 57, 58, 59, 60, 61, 62, 63, 64, 65, 66, 67, 68, 69, 70, 71, 72, 73, 74, 75, 76, 77, 78, 79, 80, 81, 82, 83, 84, 85, 86, 87, 88, 89, 90, 91, 92, 93, 94, 95, 96, 97, 98, 99, 100, 101, 102, 103, 104, 105, 106, 107, 108, 109, 110, 111, 112, 113, 114, 115, 116, 117, 118, 119, 120, 121, 122, 123, 124, 125, 126, 127, 128, 129, 130, 131, 132, 133, 134, 135, 136, 137, 138, 139, 140, 141, 142, 143, 144, 146, 145, 147, 148, 149, 150, 151, 152, 153, 154, 155, 156, 157, 158, 159, 160, 161, 162, 163, 164, 165, 166, 167, 168, 169, 170, 171, 172, 173, 174, 175, 176, 177, 178, 179, 180, 181, 182, 183, 184, 185, 186, 187, 188, 189, 190, 191, 192, 193, 194, 195, 196, 197, 198, 199, 200, 201, 202, 203, 204, 205, 206, 207, 208, 209, 210, 211, 212, 213, 214, 215]. Inevitably, such an attempt will be incomplete and every reader will have his or her own favorite reference(s) omitted. To that end we submit this document as a starting point for the motivated reader from which he/she can, perhaps, further his/her own understanding and research.

## Scaling in the study of critical behavior

Scaling has been called a pillar of modern critical phenomena" [140]. The scaling hypothesis used in the study of critical phenomena was independently developed by several scientists, including Widom, Domb, Hunter, Kadanoff, Fisher, Patashinskii and Pokrovskii (see reference [19] for an authoritative review). Much of scaling is contained in the renormalization group work of Wilson [34].

The scaling hypothesis has two categories of predictions, both of which have been verified experimentally for a variety of physical systems. The first category is a set of relations called *scaling laws*. These scaling laws relate the critical exponents $\alpha$, $\beta$ and $\gamma$ which describe, for instance, the behavior of the the specific heat ($C \sim \varepsilon^{-\alpha}$), density differences of the phases ($\rho_l - \rho_v \sim \varepsilon^{\beta}$) and isothermal compressibility ($\kappa_T \sim \varepsilon^{-\gamma}$) for fluid systems; specific heat ($C \sim \varepsilon^{-\alpha}$), magnetization ($M \sim \varepsilon^{\beta}$) and isothermal susceptibility ($\chi_T \sim \varepsilon^{-\gamma}$) for magnetic systems or the singular part of the zeroth, first and second moment of the cluster distribution percolating systems near a critical point ($\varepsilon = (T_c - T)/T_c$ for physical systems and $(p_c - p)/p_c$ for percolating systems). In all the systems mentioned here, and more, these exponents are related via the scaling law

$$\alpha + 2\beta + \gamma = 2. \qquad (1)$$

The second category is *data collapse*, which is easily demonstrated with the Ising model. We may write the equation of state as a functional relationship of the form $M = M(H, \varepsilon)$ where $H$ is the applied magnetic field. Since $M(H, \varepsilon)$ is a function of two variables, it can be graphically represented as $M$ vs $\varepsilon$ for different $H$ values. The scaling hypothesis predicts that all of these $M$ vs $\varepsilon$ curves can be "scaled" or "collapsed" onto a single curve provided that one plots a scaled $M$ ($M$ divided by $H$) to some power as a function of a scaled $\varepsilon$ ($\varepsilon$ divided by $H$ to some other power). The predictions of the scaling hypothesis are supported by a wide range of experimental work with physical systems as well as computational models [7, 14, 16, 19, 36, 37, 38, 40, 42, 46, 48, 55, 68, 79, 107, 123, 140, 159, 160, 162, 165, 181, 187, 193, 194, 198, 209, 211]. Figure 1 shows some selected examples of data collapse.

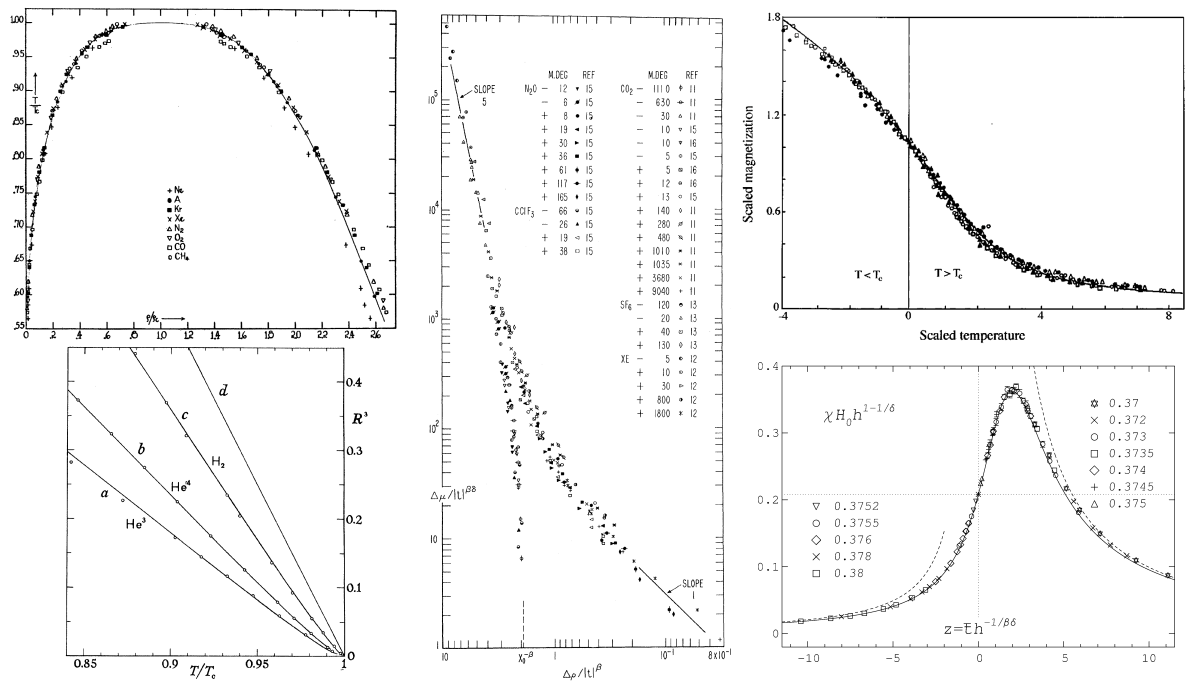

**FIGURE 1.** Examples of data collapse for various fluids and a magnetic system. Top left: the temperature $T$ divided by the critical temperature $T_c$ plotted as a function of the vapor density $\rho_v$ and liquid density $\rho_l$ normalized to the critical density $\rho_c$ [7]. Bottom left: the cube of the normalized liquid vapor density difference $R = (\rho_l - \rho_v)/\rho_c = \Delta\rho/\rho_c$ plotted as a function of the normalized temperature $T/T_c$ for "quantum" fluids (a: $He^3$, b: $He^4$ and c: $H_2$) and classical fluids (d), all fluids show scaling of the first category: $\rho_l - \rho_v \sim \varepsilon^\beta$ [14]. Center: the scaled chemical potential $|\Delta\mu|/|\varepsilon|^{\beta\delta}$ plotted as a function of the scaled density difference $|\Delta\rho|/|\varepsilon|^\beta$ in the critical region of several fluids ($CO_2$, Xe, $SF_6$, Ar, $N_2O$ and $CClF_3$) [16]. Top right: scaled experimental $MHT$ data on five different magnetic material: $CrBr_3$, EuO, Ni, YIG and $Pd_3Fe$ [140]. Bottom right: the scaled susceptibility plotted as a function of the scaled temperature for the $d = 3$ Ising model [194].

## Nuclear clusters

The success of scaling in condensed matter is impressive. However, how can this scaling be observed in nuclear reaction experiments? More specifically, how is this scaling related to the scaling of light fragment yields from nuclear reaction experiments (where direct, straightforward measurements of standard thermodynamic quantities like density, pressure, chemical potential and so on are impossible)?To see how the two are related we present a derivation of Fisher's theory in the next section.

Note: in the following text the more general term "cluster" will be used instead of "fragment" or "droplet." This is done to underscore the similarity between nuclear fragments and clusters (properly defined [25, 29, 45, 190]: all clusters in Ising calculations are so-called Coniglio-Klein clustes [45]) in systems like the Ising model and droplets of fluid (classical or quantum). We also do this to avoid the unfortunate labeling of the process of nuclear cluster production as "fragmentation" which has a specific meaning in condensed matter physics (fragments are the results of the cold fragmentation or fracturing of solid materials [76]) that may be quite different than what the nuclear multifragmentation community has in mind.

# FISHER'S THEORY AND SCALING

## Physical cluster theories

Fisher's theory is a physical cluster theory that scales [18, 19, 162, 190]. Physical cluster theories of non-ideal fluids assume that the strength of the monomer-monomer interaction is exhausted by the formation of clusters, and that the resulting clusters behave ideally (i.e. they do no interact with each other). Clusters of a given number of constituents $A$ can be characterized by their mass $m_A$, a chemical potential (per constituent) $\mu$ and a partition function $q_A(T,V)$ that depends on the temperature $T$ and volume $V$ of the fluid. Because of the ideality of the fluid of clusters, the pressure and density are straightforward to determine the pressure $p$ is

$$p = \frac{T}{V} \sum_{A=1}^{\infty} q_A(T,V) z^A \tag{2}$$

and the density $\rho$ is

$$\rho = \frac{1}{V} \sum_{A=1}^{\infty} A q_A(T,V) z^A \tag{3}$$

where $z$ is the fugacity $z = e^{\mu/T}$. The concentration of $A$ clusters is then

$$n_A(T,z) = \frac{q_A(T,V) z^A}{V}. \tag{4}$$

## Fisher's theory

Fisher's contribution to physical cluster theory was to write the partition function of a cluster in terms of the free energy of the cluster. The energetic contribution to the free energy (very recognizable to nuclear scientists) is based on the liquid drop expansion

$$E_A = E_V + E_s \tag{5}$$

where $E_V$ is the volume (or bulk) binding energy of the cluster which is taken to be

$$E_V = a_V V \simeq a'_v A. \tag{6}$$

Here $V$ is the volume of the cluster; $a_v$ is the volume energy coefficient in terms of $V$; and $a'_V$ is the volume energy coefficient in terms of $A$. The term $E_s$ is the energy loss due to the surface $s_A$ (where surface is taken as the $d-1$ measure of a cluster that exists in $d$ Euclidean dimensions) of the cluster. For clusters in $d$-dimensions this is usually taken to be

$$E_s = a_s s \simeq a'_s A^{\frac{d-1}{d}}. \tag{7}$$

Where $s$ is the surface of the cluster; $a_s$ is the surface energy coefficient in terms of $s$; and $a'_s$ is the surface energy coefficient in terms of $A^{\frac{d-1}{d}}$. Because $E_s$ is a measure of the volume energy loss due to the surface of the cluster, the surface energy coefficient is nearly equal to and opposite in sign to the volume energy coefficient: $a'_s \simeq -a'_V$. Fisher wrote the surface energy factor more generally as $E_s = a'_s A^\sigma$ where $\sigma$ is some general exponent describing the ratio of the surface to the volume of the cluster.

Fisher estimated the entropic contribution to the free energy of the cluster based on a measure of the combinatorics of the number of clusters with surface $s$ and cluster number $A$: $g_{s,A}$. Summing over the cluster number gives the total number of ways to form a cluster with a given surface $s$

$$g_s = \sum_A g_{s,A} \tag{8}$$

where the sum runs over all possible cluster numbers $A$ that result in a cluster with surface $s$. For instance, for clusters on a $d = 2$ square lattice for $s = 4$ the only possible cluster number is $A = 1$ and $g_s = 1$. The next larger possible surface is $s = 6$ and is only possible for $A = 2$ and $g_s = 2$. However, clusters with $A = 3$ and $A = 4$ can result in a

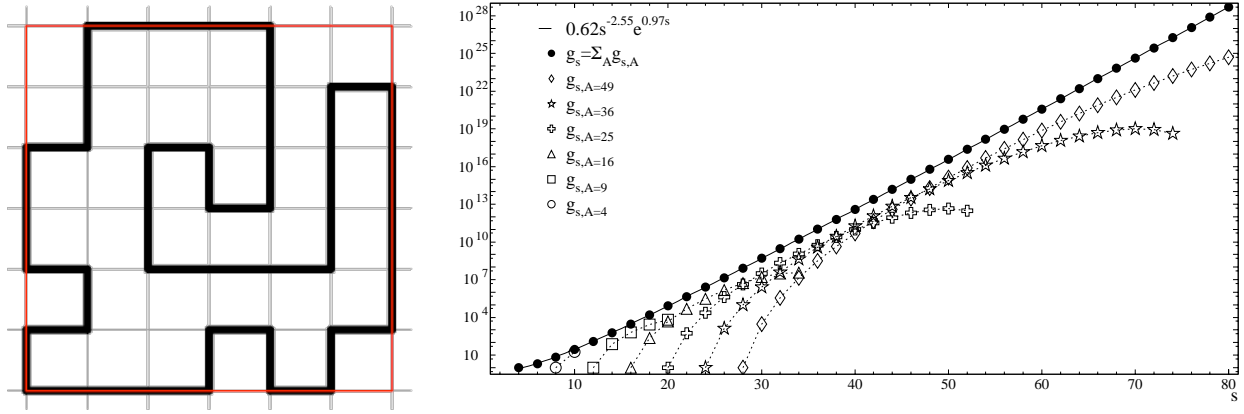

**FIGURE 2.** Left: an example of a self-avoiding polygon on the square lattice with $A=23$ and $s=40$. There are $49,157,726,494$ ways to form a cluster with this number and surface. Right: The combinatorics of self avoiding clusters on the square lattice as a function of cluster surface $s$. The open symbols show $g_{s,A}$ for $A=4,9,16,25,39,49$. The values of $g_{s,A}$ were obtained via a direct counting of the clusters [188]. The solid line shows a fit to $g_s$ using Eq. (9).

surface $s = 8$, thus the sum runs in Eq. (8) from $A = 3$ to $A = 4$ and $g_s = 7$. As the surface $s$ increases, the range of possible cluster numbers $A$ with surface $s$ increases and $g_s$ grows exponentially

$$g_s \simeq g_0 s^{-x} e^{b_s s} \tag{9}$$

where $g_0$ is an overall normalization, $b_s$ can be thought of as the limiting entropy per unit surface of a cluster. This estimate can be tested by the study (and direct counting) of the number of self-avoiding polygons on the square lattice [146, 188, 200]. An example of a self-avoiding polygons on the square lattice is shown in Fig. 2. The study of self-avoiding polygons shows that to leading order [146]

$$g_s \simeq 0.562301495 s^{-\frac{5}{2}} e^{0.97 s} \tag{10}$$

while a fit to the direct counting of self-avoiding polygons (shown in Fig. 2) gives $g_s = 0.62 s^{2.55} e^{0.97 s}$ [188, 201]. Fisher then assumed that for large clusters, over some small temperature range the most probable or mean surface of a cluster would go as

$$\bar{s} \simeq a_0 A^\sigma \tag{11}$$

so that $g_s$ could be re-written as

$$g_A \simeq g_0' A^{-\tau} e^{b_s' A^\sigma} \tag{12}$$

where $g_0' = g_0 a_0^{-x}$, $\tau = x\sigma$ and $b_s' = b_s a_0$. Which gives the entropy of a cluster as:

$$S_A = \ln g_A = \ln g_0' - \tau \ln A + b_s' A^\sigma. \tag{13}$$

The partition function of a cluster is then

$$\begin{aligned} q_A(T,V) &= V\left(\frac{2\pi m_A T}{h^2}\right)^{\frac{d}{2}} \exp\left(-\frac{E_A - TS_A}{T}\right) \\ &= V g_0' A^{-\tau} \exp\left\{\frac{\left[a_V - \frac{d}{2A} T \ln\left(\frac{h^2}{2\pi m_A T}\right)\right] A}{T}\right\} \exp\left[-\frac{(a_s' - Tb_s')A^\sigma}{T}\right]. \end{aligned} \tag{14}$$

Equation (4) then gives the cluster concentration as

$$n_A(T) = g_0' A^{-\tau} z^A \exp\left\{\frac{\left[a_V - \frac{d}{2A} T \ln\left(\frac{h^2}{2\pi m_A T}\right)\right] A}{T}\right\} \exp\left[-\frac{(a_s' - Tb_s')A^\sigma}{T}\right]$$

$$= g'_0 A^{-\tau} \exp\left\{\frac{\left[\mu + a_V - \frac{d}{2A}T \ln\left(\frac{h^2}{2\pi m_A T}\right)\right]A}{T}\right\} \exp\left[-\frac{(a'_s - Tb'_s)A^\sigma}{T}\right]. \quad (15)$$

Fisher identified the numerator of the first exponential as the distance from phase coexistence as measured by the chemical potential

$$\Delta\mu = \mu + a_V - \frac{d}{2A}T \ln\left(\frac{h^2}{2\pi m_A T}\right), \quad (16)$$

where at coexistence (or condensation) $\Delta\mu = 0$ and $\mu_{\text{coex}} = \frac{d}{2A}T \ln\left(\frac{h^2}{2\pi m_A T}\right) - a_v$.

The $(a'_s - Tb'_s)$ contribution to the surface tension vanishes at the critical point, leaving only a power law (which has been explicitly verified in computational systems [31, 42, 72, 68, 73, 74, 75, 88, 90, 91, 97, 99, 109, 117, 121, 122, 123, 124, 138, 141, 143, 150, 159, 162, 193, 198, 201, 209] and implicitly verified in a wide variety of physical fluids [26, 30]). Thus

$$T_c = \frac{a'_s}{b'_s}. \quad (17)$$

Using Eq. (16) and assuming little or no temperature dependence of $a'_s$ and $b'_s$ over the temperature range in question, then we may re-write Eq. (15) as

$$n_A(T) = g'_0 A^{-\tau} \exp\left(\frac{\Delta\mu A}{T}\right) \exp\left(-\frac{a'_s \varepsilon A^\sigma}{T}\right) \quad (18)$$

which gives the familiar expression for the cluster number concentration.

## Scaling from Fisher's theory

### Data collapse

We consider first with the second category of scaling, namely: data collapse. We start by looking at the cluster concentrations in Fisher's theory given by Eq. (18). Dividing both sides by the power law factor and the chemical potential factor then gives:

$$\frac{n_A(T)}{g'_0 A^{-\tau} \exp\left(\frac{\Delta\mu A}{T}\right)} = \exp\left(-\frac{a'_s \varepsilon A^\sigma}{T}\right). \quad (19)$$

This shows that scaling the cluster concentrations by the power law and chemical potential factors against the cluster surface free energy should collapse the data for each cluster size $A$ at each temperature $T$ to a single curve. Figure 3 shows this type of scaling and data collapse in percolation [199] and Ising model cluster yields [198]

### Scaling relations

To arrive at the first category of scaling from Fisher's theory, we combine the general equations for pressure and density for physical cluster theories, equations (2) and (3), with Fisher's estimate of the cluster partition function, Eq. (15) giving

$$p = T \sum_{A=1}^{\infty} g'_0 A^{-\tau} \exp\left(\frac{\Delta\mu A}{T}\right) \exp\left(-\frac{a'_s \varepsilon A^\sigma}{T}\right) \text{ and } \rho = \sum_{A=1}^{\infty} g'_0 A^{1-\tau} \exp\left(\frac{\Delta\mu A}{T}\right) \exp\left(-\frac{a'_s \varepsilon A^\sigma}{T}\right). \quad (20)$$

Along the coexistence line, i.e. $\Delta\mu = 0$, we have

$$p_{\text{coex}} = T \sum_{A=1}^{\infty} g'_0 A^{-\tau} \exp\left(-\frac{a'_s \varepsilon A^\sigma}{T}\right) \text{ and } \rho_{\text{coex}} = \sum_{A=1}^{\infty} g'_0 A^{1-\tau} \exp\left(-\frac{a'_s \varepsilon A^\sigma}{T}\right). \quad (21)$$

At the critical point we have

$$p_c = T_c \sum_{A=1}^{\infty} g'_0 A^{-\tau} \text{ and } \rho_c = \sum_{A=1}^{\infty} g'_0 A^{1-\tau}. \tag{22}$$

Taking the ratios of equations (21) to (22) gives the reduced pressure $p_{\text{coex}}/p_c$ and reduced density $\rho_{\text{coex}}/\rho_c$

$$\frac{p_{\text{coex}}}{p_c} = \frac{T \sum_{A=1}^{\infty} A^{-\tau} \exp\left(-\frac{a'_s \varepsilon A^\sigma}{T}\right)}{T_c \sum_{A=1}^{\infty} A^{-\tau}} \text{ and } \frac{\rho_{\text{coex}}}{\rho_c} = \frac{\sum_{A=1}^{\infty} A^{1-\tau} \exp\left(-\frac{a'_s \varepsilon A^\sigma}{T}\right)}{T_c \sum_{A=1}^{\infty} A^{1-\tau}} \tag{23}$$

which has the advantage of being free of the constant $g'_0$. In order to further test the results above, we determine the magnetization $M$ of the $d = 3$ Ising model using Eq. (23) and recalling that the magnetization per lattice site is simply:

$$M = 1 - \frac{\rho}{\rho_c}. \tag{24}$$

Using the values of $\sigma$, $\tau$, $c_0$ and $T_c$ determined from fitting clusters on the $d = 3$ Ising lattice shown in Fig. 3 [198] in Eq. (23), Eq. (24) gives one branch of the magnetization curve, the branch for $M > 0$. Since the magnetization is symmetric about the origin, the points for $M < 0$ are reflections of the points for $M > 0$. The results are shown as the open circles in the bottom right plot of Fig. 3. These results compare well with a parametrization for $M(T)$ [198] (used as a "benchmark") shown as a solid line in the bottom right plot of Fig. 3. Better agreement with the $M(T)$ parameterization is found when the values of $\sigma = 0.63946 \pm 0.0008$, $\tau = 2.209 \pm 0.006$ (from the scaling relations in Fisher's theory developed below and values of $\beta = 0.32653 \pm 0.00010$ and $\gamma = 1.2373 \pm 0.002$ [184]), $a'_s = 12$ and $T_c = 4.51152 \pm 0.00004$ were used. Nearly perfect results were observed when $a'_s$ was "tuned" to 16 and the more precise value of $T_c$ and the scaling relation exponent values were used. The agreement between the magnetization values calculated via the sum in Eq. (24) and the $M(T)$ parameterization for $0 < T < T_c$ suggest that the ideal gas assumptions in Fisher's theory allow for an accurate description of the system even up to densities as high as $\rho_c$.

By combining equations (21) and (22) we can arrive at the scaling relations as follows:

$$\frac{\rho_c - \rho_{\text{coex}}}{\rho_c} = \frac{g'_0}{\rho_c} \sum_{A=1}^{\infty} A^{1-\tau} \left[1 - \exp\left(-\frac{a'_s \varepsilon A^\sigma}{T}\right)\right] \simeq \frac{g'_0}{\rho_c \sigma} \Gamma\left(-\frac{\tau-2}{\sigma}\right) \left(\frac{a'_s}{T_c}\right)^{\frac{\tau-2}{\sigma}} \varepsilon^{\frac{\tau-2}{\sigma}} = B\varepsilon^\beta \tag{25}$$

since as $T \to T_c$ large values of $A$ give the dominant contribution to the above sum and the sum may be replaced by an integral ($\int_0^\infty y^{x-1} e^{-y} dy = \Gamma(x)$) [39]. Here $\beta = \frac{\tau-2}{\sigma}$. This leads directly to the familiar relation $\rho_l - \rho_v \sim \varepsilon^\beta$.

Similarly, one finds that along the coexistence line the specific heat at constant volume is [18, 190]

$$C_V = T^2 \left.\frac{\partial^2 \frac{p_{\text{coex}} V}{T}}{\partial T^2}\right|_V \sim \varepsilon^{2-\frac{\tau-1}{\sigma}} \sim \varepsilon^{-\alpha} \tag{26}$$

thus $\alpha = 2 - \frac{\tau-1}{\sigma}$.

Finally, the isothermal compressibility can be found to be [162]

$$\kappa_T = \frac{1}{\rho} \left.\frac{\partial \rho}{\partial p}\right|_T \sim \varepsilon^{\frac{\tau-3}{\sigma}} \sim \varepsilon^{-\gamma} \tag{27}$$

thus $\gamma = \frac{3-\tau}{\sigma}$.

The three examples above show how Fisher's theory leads to the power laws that describe the behavior of a system near its critical point. Putting the equations defining $\alpha$, $\beta$ and $\gamma$ together recovers the scaling law $\alpha + 2\beta + \gamma = 2$ and illustrates that (aside from so-called "hyperscaling") there are only two independent exponents ($\sigma$ and $\tau$ in Fisher's theory) from which all others are recovered.

## Caveats

Before proceeding further, we must study the implications and assumptions inherent to Fisher's theory and any problems, inconsistencies or discrepancies that arise because of them.

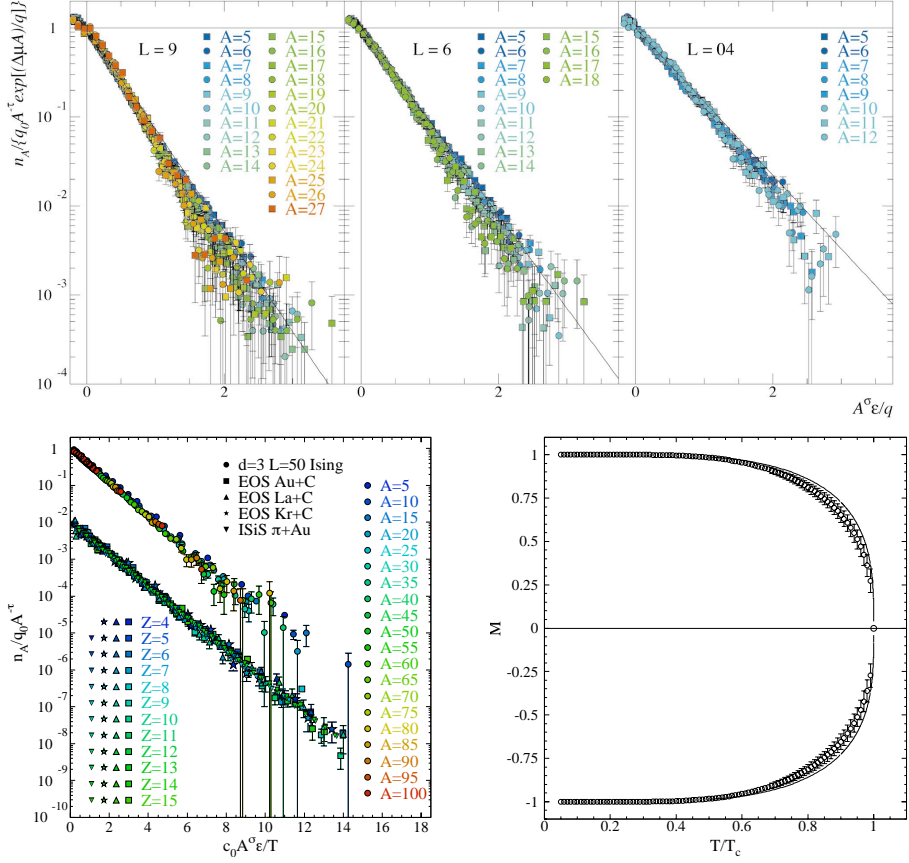

**FIGURE 3.** Top three plots: the scaling of Eq. (19) for $d = 3$ bond breaking percolation on the simple cubic lattice of side $L = 9$, 6 and 4 ($q$ is the bond breaking probability) [193]. Bottom left plot: the scaling of Eq. (19) for $d = 3$ Ising model on the simple cubic lattice of side $L = 50$ [198]. Bottom right plot: the magnetization as a function of reduced temperature. The open circles show the magentization predicted via Fisher's theory (see text) and the solid line shows a parameterization for the magnetization.

*Surface tension*

The first implication is that a cluster's surface free energy is linear in $\varepsilon$. This implication appears in Fisher's work only when deriving the singular behavior of quantities near critical point (e.g. the isothermal compressibility and the liquid-vapor density difference) [18] and does not appear explicitly in the cluster concentrations until other work with Fisher's theory [31, 35, 39, 42].

To see the linearity in the surface free energy most clearly, we can recast Fisher's theory solely in terms of the surface area of a cluster $s$, the concentration is the product of the combinatorial factor and a Boltzmann factor that depends on the surface energy:

$$n_s(T) \propto g_s \exp\left(-\frac{a_s s}{T}\right). \tag{28}$$

Following the arguments put forward in the preceding section and using Eq. (9) gives

$$n_s(T) \propto s^{-x} \exp\left(-\frac{a_s \varepsilon s}{T}\right), \tag{29}$$

suggesting that the surface tension coefficient of a cluster is $a_s \varepsilon$.

However, it has long been known empirically [2] that the surface tension of macroscopic fluids $\Gamma$ (the surface free energy per unit area) is not linear in $\varepsilon$. In fact, to lowest order, as $T \to T_c$ [15, 19, 20]

$$\Gamma = \Gamma_0 \varepsilon^{(d-1)\nu} \tag{30}$$

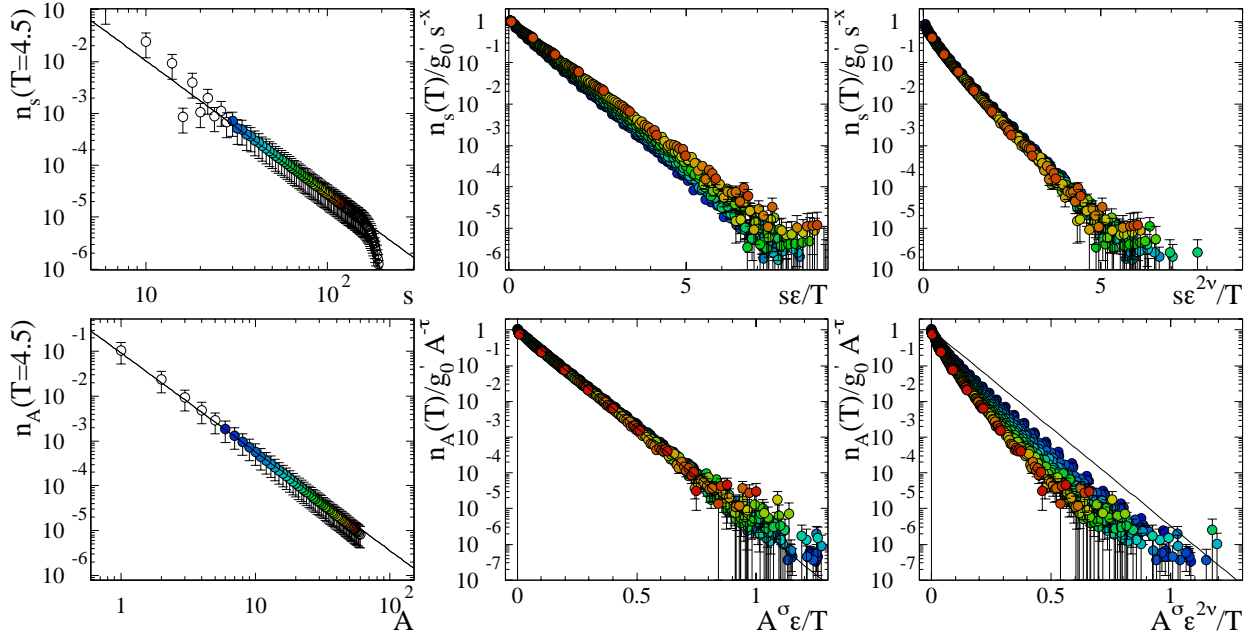

**FIGURE 4.** The scaling of cluster concentrations $n_s(T)$ from $d=3$ Ising calculations on the simple cubic lattce [211]. Top (from left to right): The cluster concentration at $T = 0.997T_c$ in terms of cluster surface $s$ (the solid line shows the cluster concentration at the critical point which is empirically estimated to go as $n_s(T_c) = 3.85s^{-2.569}$); the scaling of $n_s(T)$ according to Fisher's theory with a surface free energy which varies linearly in $\varepsilon$; and the scaling of $n_s(T)$ according to Fisher's theory with a surface free energy which varies as $\varepsilon^{2\nu}$. Bottom (from left to right): The cluster concentration at $T = 0.997T_c$ in terms of cluster number $A$ (the solid line shows the cluster concentration at the critical point which goes as $n_A(T_c) = 0.093s^{-2.209}$ [198]); the scaling of $n_A(T)$ according to Fisher's theory with a surface free energy which varies linearly in $\varepsilon$; and the scaling of $n_A(T)$ according to Fisher's theory with a surface free energy which varies as $\varepsilon^{2\nu}$ (the solid line shows the value of the surface free energy coefficient $a'_s = 12.63 \pm 0.04$ [211]). Colors give the surface or number of the cluster as can be seen in the left most plots. No fitting has been done in any of the scaling plots of this figure.

where $\nu$ is the critical exponent that also describes the divergence of the correlation length near the critical point and is related to other exponents through the hyper-scaling relation [15, 19]

$$d\nu = \gamma + 2\beta = 2 - \alpha - \frac{\tau-1}{\sigma}. \qquad (31)$$

Studies of the $d=3$ Ising model indicate that the surface tension is sensitive to higher order terms (H.O.T.s)

$$\Gamma = a_s \varepsilon^{2\nu}\left(1 + a_\theta \varepsilon^\theta + a_1 \varepsilon\right) \qquad (32)$$

with $a_s = 1.55 \pm 0.05$, $a_\theta = -0.41 \pm 0.05$, $\theta = 0.51$ and $a_1 = 1.2 \pm 0.1$ [126].

Futhermore, it has long been known that the surface tension of a cluster $\Gamma(A,T)$ may differ from surface tension of a macroscopic fluid $\Gamma$ [8, 210]

$$\Gamma(A,T) = \Gamma\left(1 - \frac{2\delta_r}{r_A}\right) \qquad (33)$$

where the Tolman length $\delta_r$ is independent of cluster sizes and $r_A$ is the radius of the cluster. However, this affects only the magnitude of the surface tension, whereas the temperature dependence of the surface tension remains the same for clusters and the macroscopic fluid.

The top panels of Fig. 4 shows that the concentrations of clusters as a function of their surface $n_s(T)$ in $d=3$ Ising calcutaions [211] are poorly described by Eq. (29) and better described by

$$n_s(T) \propto s^{-x} \exp\left(-\frac{a_s \varepsilon^{2\nu} s}{T}\right), \qquad (34)$$

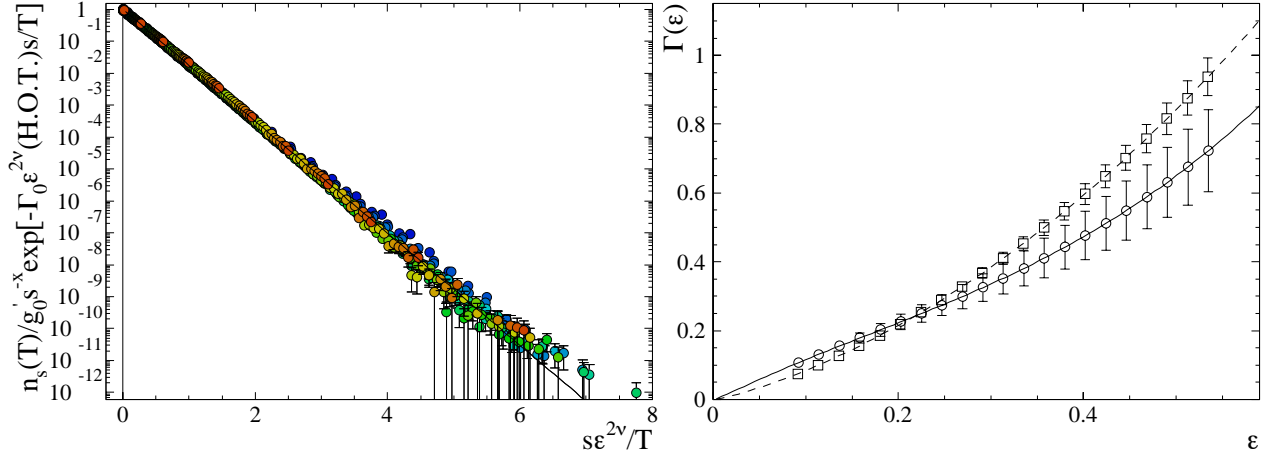

**FIGURE 5.** Left: the scaling of cluster concentrations $n_s(T)$ from $d = 3$ Ising calculations on the simple cubic lattce [211] according to Eq. (35). Over 700 points are collapsed to a single curve with the fit parameters: $T_c = 4.516 \pm 0.004$, $g'_0 = 4.3 \pm 0.2$, $x = 2.58 \pm 0.01$, $a_s = 4.04 \pm 0.09$, $a_\theta = -2.11 \pm 0.04$, $a_1 = 1.73 \pm 0.06$ and $\delta_s = -1.7 \pm 0.1$. Colors give the surface of the cluster as in Fig. 4. Right: a comparison of the behavior of the surface tension from the fit to the left (open circles) and the surface tension determined in reference [126] (open squares), while the parameters are different, the overall behavior is similar over the temperatures considered: $2.5 \leq T \leq 4.5$.

with $\nu = 0.6299 \pm 0.0002$ set to its $d = 3$ Ising value and $T_c = 4.51152 \pm 0.00004$ set to its value for the simple cubic lattice [184], the exponent $x$ is determined empirically from a power law fit to the cluster concentrations at $T = T_c$. Thus when counting clusters in terms of surface $s$ a surface tension that varies as $\varepsilon^{2\nu}$ is needed to describe the cluster distrubtions. The top panels of Fig. 4 show, empirically, the same is not true when scaling in terms of cluster number $A$: Eq. (15), where a surface free energy linear in $\varepsilon$ gives a better description than the form $\varepsilon^{2\nu}$.

As an aside we note that the data collapse shown in Fig. 4 can be improved by taking into account the higher order terms in Eq. (32) and the cluster size effects in Eq. (33) the cluster concentration of Eq. (34) to give

$$n_s(T) = g'_0 s^{-x} \exp\left[-\frac{s a_s \varepsilon^{2\nu}\left(1 + a_\theta \varepsilon^\theta + a_1 \varepsilon\right)\left(1 - \frac{2\delta_s}{\sqrt{s}}\right)}{T}\right]. \tag{35}$$

Figure 5 shows the results for the scaled cluster concentrations and the surface tension.

To understand why the surface free energy in terms of cluster number $A$ is more accurately described by a surface free energy linear in $\varepsilon$ we look in more detail at the change in describing the cluster concentrations in terms of cluster number $A$ rather than the cluster's surface $s$. Working in $d = 3$ for the sake of illustration and assuming that the clusters are spherical (which will be tested below) we have the cluster's surface as:

$$s = 4\pi r_A^2 \tag{36}$$

where $r_A$ is the radius of the cluster in question. The cluster's volume is

$$V = \frac{4}{3}\pi r_A^3. \tag{37}$$

Using the density of the cluster $\rho = A/V$ shows that

$$s = 4\pi \left(\frac{3}{4\pi\rho}\right)^{2/3} A^{2/3}. \tag{38}$$

If we treat each cluster in the vapor as a small drop of liquid, then the pertinent density is the density of the liquid $\rho_l$ and

$$s = 4\pi \left(\frac{3}{4\pi\rho_l}\right)^{2/3} A^{2/3}. \tag{39}$$

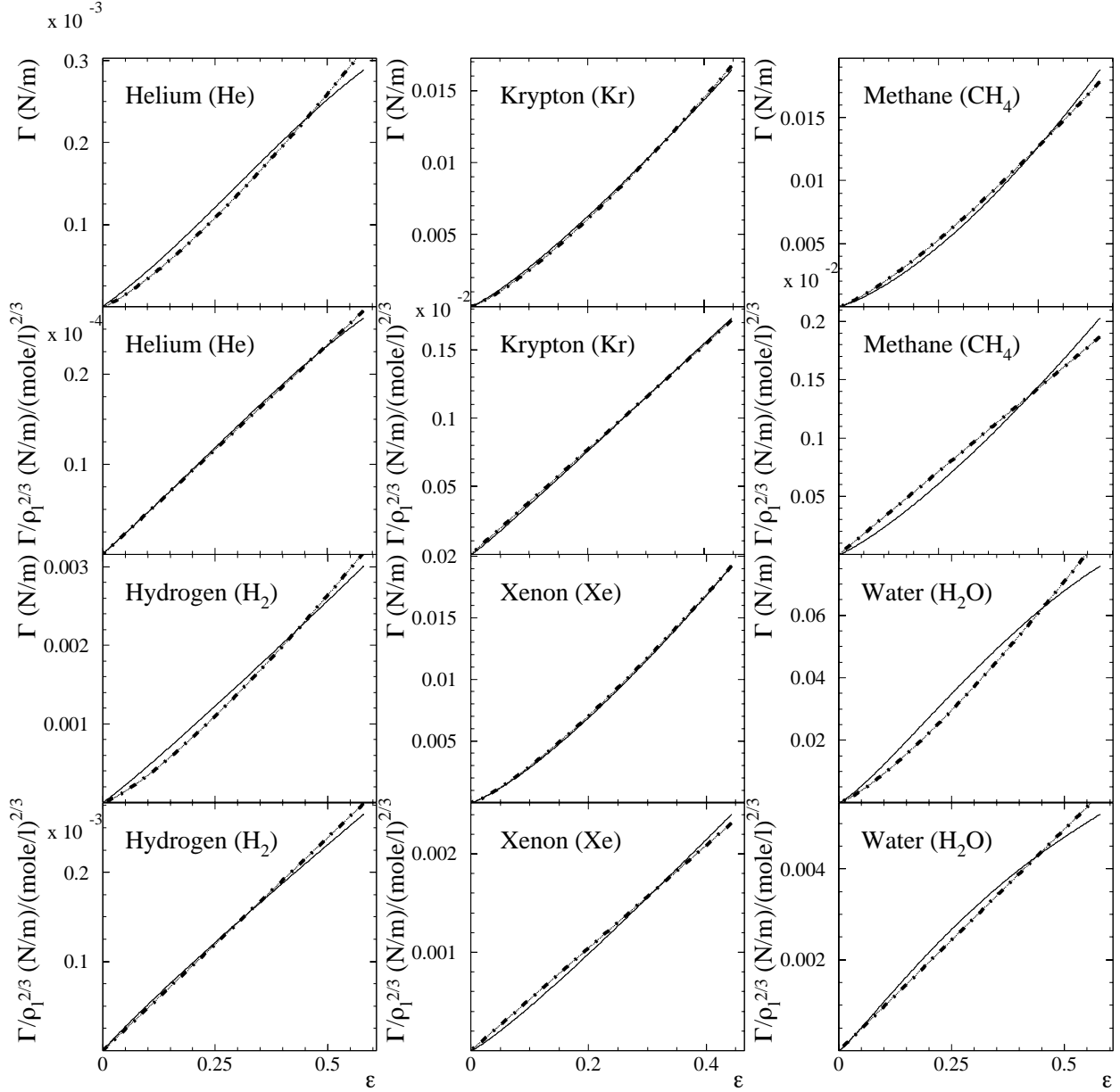

**FIGURE 6.** The surface tension $\Gamma$ (in $N/m$) in terms of cluster surface $s$ (first and third rows) and the surface tension $\Gamma/\rho_l^{2/3}$ (in $(N/m)/(mole/l)^{2/3}$) in terms of cluster number $A$ (second and fourth rows)) as a function of $\varepsilon = (T_c - T)T_c$ for: "quantum" fluids (hydrogen and helium), noble gases (krypton and xenon) and more complex fluids (methane and water). The thin solid lins show data points [213] and the heavy dashed-dotted lines over the thin line show fits to $\Gamma = \Gamma_0 \varepsilon^{2\nu}$ and $\Gamma/\rho_l^{2/3} = \Gamma'_0 \varepsilon$.

The surface free energy is

$$F_s = \Gamma_0 \varepsilon^{2\nu} s = \Gamma_0 \varepsilon^{2\nu} 4\pi \left(\frac{3}{4\pi\rho_l}\right)^{2/3} A^{2/3} = \frac{\Gamma}{\rho_l^{2/3}} 4\pi \left(\frac{3}{4\pi}\right)^{2/3} A^{2/3}. \qquad (40)$$

Thus when writing the surface free energy in terms of cluster number $A^{2/3}$ the surface tension $\Gamma$ is effectively modified by $\rho_l^{2/3}$ which varies as $\varepsilon^\beta$ as $T \to T_c$.

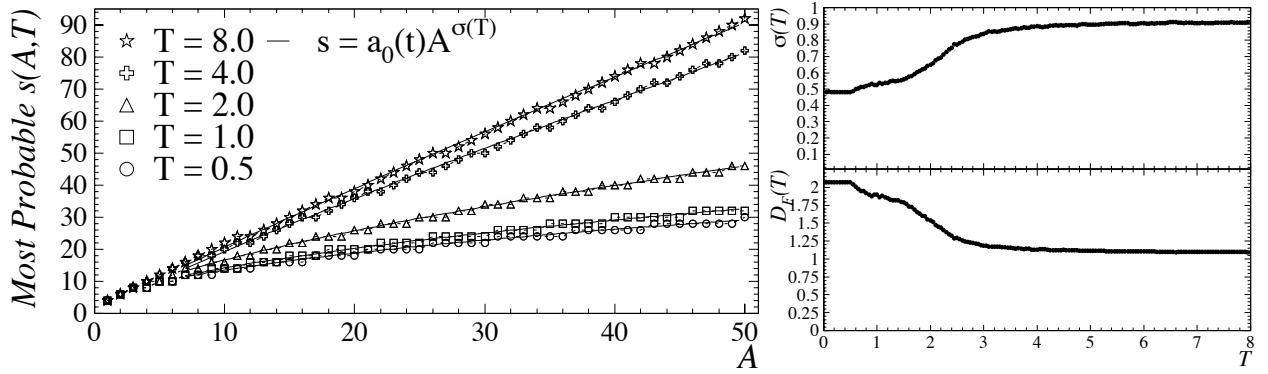

**FIGURE 7.** Left: the most probable surface as a function of $A$ and $T$. Right: top: the effective surface to volume exponent $\sigma$ as a function of temperature; bottom: the fractal dimension $D_F = 1/\sigma$ [42] as a function of temperature, see references [68, 201] for details.

To see the effect of the change from cluster surface to cluster number we look at the surface tension for a selection of real fluids shown in Fig. 6. For a broad range of fluids as $\varepsilon \to 0$ Eq. (30) describes the behavior of the surface tension. However, the ratio of $\Gamma/\rho_l^{2/3}$ is adequately described by

$$\frac{\Gamma}{\rho_l^{2/3}} = \Gamma_0' \, \varepsilon. \tag{41}$$

Figure 6 shows empirically that the modification of $\Gamma$ by $\rho_l^{2/3}$ effectively changes the power law variation from $\varepsilon^{2\nu}$ to $\varepsilon^1$.

This change can also be seen analytically by looking at the effective exponent $\nu_{\text{eff}}$ of $\varepsilon$ of $F_s$ in Eq. (40):

$$\nu_{\text{eff}} = \frac{\partial \ln F_s}{\partial \ln \varepsilon} = 2\nu - \frac{2}{3} \frac{d_1 \varepsilon + d_\beta \beta \varepsilon^\beta}{1 + d_1 \varepsilon + d_\beta \varepsilon^\beta} \tag{42}$$

where the parameterization

$$\rho_l = \rho_c \left(1 + d_1 \varepsilon + d_\beta \varepsilon^\beta\right) \tag{43}$$

was used for the density of the liquid [7]. In the limit of $\varepsilon \to 0$: $\nu_{\text{eff}} \to 2\nu$ but in the limit of $\varepsilon \to 1$: $\nu_{\text{eff}} \to 1$ for systems in the $d = 3$ Ising class. Thus, it is clear that the translation from cluster surface to cluster number causes the surface free energy to vary, approximately, linearly in $\varepsilon$ away from the critical point.

*Cluster shape*

Next we can examine Fisher's assumptions on the most probable or mean surface $\bar{s}$ of a cluster. We may do this by using the combinatorics of self-avoiding polygons and noting that, at phase coexistence where $\Delta\mu = 0$, Eq. (44) is the product of the combinatorial factor and a Boltzmann factor that depends on the surface energy:

$$n_{s,A}(T) \propto g_{s,A} \exp\left(-\frac{a_s s}{T}\right) \tag{44}$$

where now we write the cluster concentrations explicitly in terms of both cluster number $A$ and cluster surface $s$ [199, 201]. The mean surface of a cluster is then just

$$\bar{s} = \frac{\sum_{A=1}^{\infty} s n_{s,A}(T)}{\sum_{A=1}^{\infty} n_{s,A}(T)}. \tag{45}$$

Using the direct counting of $g_{s,A}$ (see Fig. 2) and setting (as in the Ising model) $a_s = 2$ (thus $T_c \simeq \frac{2}{0.97} = 2.06$) we can determine the most probable surface of a cluster of $A$ constituents at temperature $T$. Fitting $\bar{s}(A)$ with $a_0 A^\sigma$ letting $a_0$ and $\sigma$ be free parameters we can study Fisher's assumption. Figure 7 shows that at low temperatures $\sigma \simeq 0.5$ as one would expect for a $d = 2$ system. As the temperature increases the value of $\sigma$ increases. At $T = T_c \simeq 2.06$, $\sigma \simeq 0.65$, a change of 30% from the $T = 0$ value of $\sigma$. Thus, Fisher's implicit assumption that $\sigma$ is a constant is only accurate to the 30% level in this example. Looking at the accpeted value of $\sigma = 8/15$ from the $d = 2$ Ising model [209] and comparing it to the expected $T = 0$ value of $\sigma = 1/2$ shows this assumption to be accurate to the 6.67% level for $0 \leq T \leq T_c$. Looking at the accpeted value of $\sigma = 0.63946 \pm 0.0008$ from the $d = 3$ Ising model [198] and comparing it to the expected $T = 0$ value of $\sigma = 2/3$ shows this assumption to be good to the 4.08% level for $0 \leq T \leq T_c$. When the temperature is restricted to a very small range around $T \sim T_c$ this assumption is quite good.

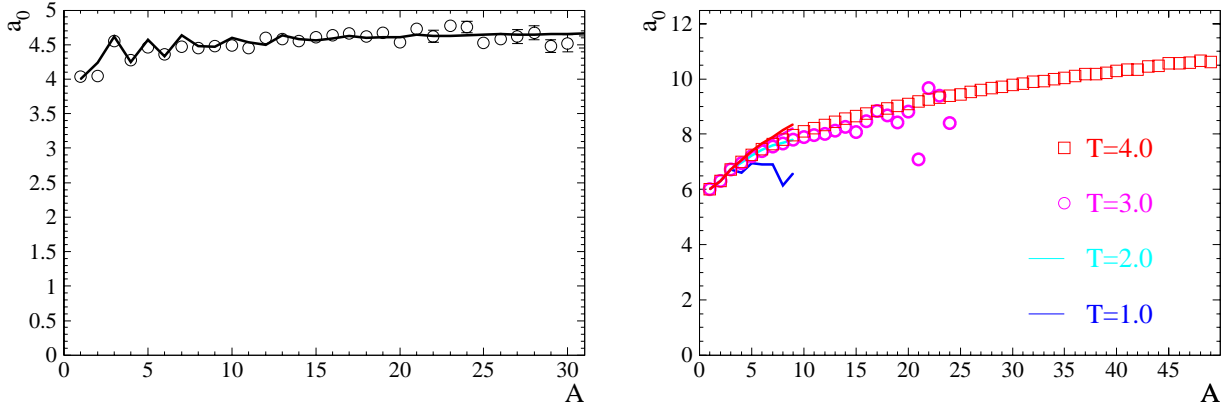

**FIGURE 8.** Left: the geometrical pre-factor for the mean surface to cluster size relation $a_0$ as a function of cluster size $A$. The solid line shows the results for Eq. (46), open circles show $a_0$ for $d = 2$ Ising clusters [199]. Right: the geometrical pre-factor for the mean surface to cluster size relation $a_0$ as a function of cluster size $A$. The solid line shows the results for Eq. (46), open circles show $a_0$ for $d = 3$ Ising clusters and the colors show different temperatures. Error bars are smaller than the points.

Another possible problem with this assumption is not only the dependence of $\sigma$ on temperature, but the dependence of $a_0$ on temperature and cluster size $A$. Fisher implicitly assumed that for $A \to \infty$ $a_0$ is some constant. Using Eq. (45) with the $g_{s,A}$ of self-avoiding polygons we can test this assumption, by examining

$$a_0 = A^{-\sigma}\bar{s} = A^{-\sigma}\frac{\sum_{A=1}^{\infty} s n_{s,A}(T)}{\sum_{A=1}^{\infty} n_{s,A}(T)}. \tag{46}$$

In this example $\sigma = 8/15$ is taken from the $d = 2$ Ising model and $T = 1 \simeq T_c/2$. Figure 8 shows the results. For $A < 10$ the value of $a_0$ clearly shows "shell effects" that cause fluctuations on the order of 10% of the limiting value of $a_0$. For $A \geq 10$ the shell effects diminish and the limiting value of $a_0 \simeq 4.6$ is reached. Thus in this example Fisher's assumption holds for $A \geq 10$ [199].

Figure 8 also shows the results from a direct counting of $d = 3$ self-avoiding polyhedra [203] and clusters from the $d = 3$ simple cubic Ising lattice [211]. The $g_{s,A}$ for the self-avoiding polyhedra has been directly counted up to $A = 9$, counting for $A \geq 10$ is prohibitively time consuming on today's computers. However, the dependence of $a_0$ on cluster size and temperature can be investigated just as in the case of the $d = 2$ polygons (using $\sigma = 0.63946 \pm 0.0008$ and $a_s = 2$, which holds for the $d = 3$ Ising model as well). We see that for the lowest temperature ($T = 1$, as compared to the $d = 3$ Ising model $T_c = 4.51152 \pm 0.00004$ [184]) the shell effects are evident: for perfect cubes $A = 1$ and $A = 9$ $a_0 = 6$ as expected. As the temperature increases the shell effects are washed out and $a_0$ shows a steady rise with $A$. The steady rise of $a_0$ with $A$ could indicate that $\sigma \geq 0.63946 \pm 0.0008$ (which violates the first category of scaling as will be seen below) or that the limiting behavior Fisher assumed does not set in until $A \gg 50$. In either case, it seems this assumption is poorer for $d = 3$ than for $d = 2$.

## Excluded volume

The final entry into this section discussing caveats to Fisher's theory is the effect of the extended, non-zero volume of real, physical clusters, the so-called excluded volume. Fisher's theory, like any physical cluster theory, assumes that the clusters are point clusters and have no volume. Obviously this is not the case for real fluids, so how well does Fisher's theory do in describing real clusters which have non-zero volume [41]? We have already seen in Fig. 3 that Fisher's theory collapses the cluster concentrations of computer models quite well when the parameters (exponents, critical temperature, surface energy coefficient) are allowed to vary; the values returned for these parameters from the fitting procedures usually agree well with expected values [193, 198] (with the exception of $\sigma$ for the $d = 3$ Ising model, though that discrepancy may be the result of using clusters that are too small, see Fig. 8 and the discussion above).

In the case of physical fluids the effects of the extended volume of clusters at the critical point can be studied by realizing that Fisher's theory gives the compressibility factor $C_F$ as the ratio of two Riemann $\zeta$ functions

$$C_F = \frac{p_c}{T_c \rho_c} = \frac{\sum_{A=1}^{\infty} A^{-\tau}}{\sum_{A=1}^{\infty} A^{1-\tau}} = \frac{\zeta(\tau)}{\zeta(\tau-1)} = 0.277 \pm 0.004. \tag{47}$$

The top panel of the left most plot of Fig. 9 shows the value of $C_F$ as a function of $T_c$ for 30 fluids from helium through water. When the compressibility factor for real fluids (e.g. He$^4$, Ne, ethane, acetylene, CH$_3$CH, C$_2$H$_5$Cl, etc.) was analyzed it was found that $\tau = 2.202 \pm 0.004$ which is to be expected for $d = 3$ systems [26]. This result indicates that for real fluids the value of $\tau$ is not greatly affected by the finite size of the clusters. Furthermore, an analysis of the "excluded-volume" effect and Fisher's theory later showed that the scaling laws (e.g. $\alpha + 2\beta + \gamma = 2$) were unchanged [28].

More insight to this can be gained by examining an attempt that was made to explicitly correct Fisher's theory [18] for the effect of the cluster's extended volume by the adding of a Boltzmann factor with $bpA$ to Eq. (21). Thus the pressure is given by [167]

$$p = T \sum_{A=1}^{\infty} g'_0 A^{-\tau} \exp\left(\frac{\Delta \mu A}{T}\right) \exp\left(-\frac{a'_s \varepsilon A^{\sigma}}{T}\right) \exp\left(-\frac{bpA}{T}\right) \tag{48}$$

where $b$ is the one-particle volume as in the van der Waals fluid [1] (and thus $bA$ is the volume of a cluster of $A$ particles). The vapor density is then found from

$$\rho = \frac{\partial p}{\partial \Delta \mu} = \frac{g'_0 \sum_{A=1}^{\infty} A^{1-\tau} \exp\left(\frac{\Delta \mu A - a'_s A^{\sigma} \varepsilon - bpA}{T}\right)}{1 + b g'_0 \sum_{A=1}^{\infty} A^{1-\tau} \exp\left(\frac{\Delta \mu A - a'_s A^{\sigma} \varepsilon - bpA}{T}\right)}. \tag{49}$$

At the critical point: $\varepsilon = 0$, $\Delta \mu - b p_c = 0$ so that the pressure and density at the critical point are given by

$$p_c = T_c g'_0 \sum_{A=1}^{\infty} A^{-\tau} \text{ and } \rho_c = \frac{g'_0 \sum_{A=1}^{\infty} A^{1-\tau}}{1 + b g'_0 \sum_{A=1}^{\infty} A^{1-\tau}}. \tag{50}$$

The sums in Eq. (50) are Riemann $\zeta$-functions and easily calculated. For $\tau = 2.209$ [26] then

$$p_c = T_c g'_0 1.48488211 \text{ and } \rho_c = \frac{g'_0 5.37690859}{1 + b g'_0 5.37690859}. \tag{51}$$

Combining equations (47) and (51) yields an estimate for the one-particle volume $b = -0.001 \pm 0.003/g'_0$. The middle panel of the left most plot of Fig. 9 shows that for fluids $g'_0 \approx 2$ mole/$l$ (determined from the empirical value of $p_c$ and Eq. (51)) so that $b \approx (-6 \pm 7) \times 10^{-24}$ cm$^3$ per particle (bottom panel of the left most plot of Fig. 9). Typical atomic radii for fluids are on the order of 100pm so the volume of a single particle in the fluid is approximately $V_{\text{particle}} \approx 4.2 \times 10^{-21}$ cm$^3$.

Since $b \approx V_{\text{particle}}/700$ the effects of excluded volume on Fisher's theory at the critical point are negligible. This is obvious when one examines the value of $C_F$ given by Fisher's theory assuming $b = 0$ and putting $\tau = 2.209$ into the Riemann $\zeta$-functions in Eq. (47): $C_F = 0.276159$ which agrees, to within error bars, with the empirical value.

Furthermore, $\rho_c$ is the highest density at which Fisher's theory is applicable, thus this exercise has provided us with empirical evidence that Fisher's theory is unaffected by one-particle volume effects from $0 \leq T \leq T_c$.

If the exponents, scaling laws and compressibility factor are unaffected by the extended volume of clusters, then what are the effects of the non-zero volume of the clusters? To answer this question we turn our attention back to the self-avoiding polygons [200]. Figure 8 shows that using the directly counted combinatorics $g_{s,A}$ we were able to reproduce the behavior of clusters from the $d=2$ lattice gas (Ising) model on a square lattice, up to a point. The critical temperature predicted by the self-avoiding polygons $T_c = 2.06$ is approximately 10% below Onsager's analytically determined value $T_c = 2.26915\ldots$ [6].

To improve the above estimate of $T_c$, at coexistence, we think of an initial configuration of a cluster with $A_0 \to \infty$ constituents and surface $s_0$ and a final state of a cluster of $A$ constituents and surface $s$ and its complement: a cluster of $A_c = A_0 - A$ constituents and surface $s_c$. This assumes stochastic cluster formation and is supported by the cluster's Poissonian nature [198]. Now the free energy of cluster formation is

$$\Delta G = \Delta E - T\Delta S + p_{\text{coex}}\Delta V = a_V\left[A + (A_0 - A) - A_0\right] + a_s(s + s_c - s_0) - T\left(\ln g_{s,A} + \ln g_{s_c,A_c} - \ln g_{s_0,A_0}\right) + p\Delta V \quad (52)$$

$\Delta V$ is the volume change between the initial and final configurations. All terms $\propto A$ cancel. In the limit $A_0 \to \infty$, $s_c \approx s_0$ and $\ln g_{s_c} \approx \ln g_{s_0}$ leaving only the cluster's contribution to the $\Delta G$. The volume change for the lattice gas is

$$\Delta V = a_1\left[A + (A_0 - A) - A_0\right] + l(s + s_c - s_0) \quad (53)$$

where $a_1$ is the geomertrical prefactor relating the cluster volume to the cluster number $A$ and $l$ is the interaction range between two constituents, one spacing on a lattice: $l = 1$. The second term of Eq. (53) arises from the fact that no two clusters can come within a distance $l$ of each other and be considered two clusters, thus each cluster has a volume $ls$ surrounding it which is excluded to all other clusters.

In the $A_0 \to \infty$ limit the first term of Eq. (53) vanishes, thus the inclusion of the Boltzmann factor with $bpA$ in Eq. (48) is incorrect. The second term of Eq. (53) depends only on the cluster's surface. Writing the partition function for a cluster as $q_s(V,T) \sim \exp(-\Delta G/T)$ [39] and now including the excluded volume factor from Eq. (53) gives

$$n_s(T) \sim g_s \exp\left(-\frac{a_s s}{T}\right)\exp\left(-\frac{2p_{\text{coex}} ls}{T}\right) \sim g_0 s^{-x} \exp\left[-\frac{s(a_s - Tb_s + 2p_{\text{coex}}l)}{T}\right]. \quad (54)$$

The factor of two in the final Boltzmann factor arises when thinking in terms of transition state theory with the initial state being a liquid drop with no bubbles surrounded by a vapor and the final state being the liquid drop with a bubble the same size as the new cluster in the vapor surrounding the liquid drop. Both the bubble in the condensed phase and the cluster in the dilute phase have the associated excluded volume contribution of $ls$.

Just as above, the $(a_s - Tb_s + 2p_{\text{coex}}l)$ portion of the surface free energy vanishes at the critical point so

$$T_c = \frac{a_s + 2p_c l}{b_s} = \frac{a_s}{b_s} + \frac{2p_c l}{b_s}. \quad (55)$$

The first term in Eq. (55) can be thought of as the "ideal" critical temperature and the second term can be thought of as the correction that arises due to the non-zero volume of the cluster. Working at the critical point with $p_c \approx 0.11$ for the $d=2$ lattice gas (Ising) model, Eq. (55) gives $T_c = 2.29$, within 1% of the Onsager value [6].

The $\varepsilon$ version of Fisher's theory can be recovered by letting:

$$a'_s = a_s + 2p_c l \quad (56)$$

so that

$$a_s - Tb_s + 2p_c = a'_s - Tb_s = a'_s\varepsilon \quad (57)$$

with $b_s = a'_s/T_c$.

Equation (54) also provides a good description of Ising cluster yields. The middle plot of Fig. 9 shows the lattice gas (Ising) yields ($n_A(T) = \sum_s n_{s,A}(T)$) of a two dimensional square lattice of side $L = 80$ and the predictions of Eq. (18) and (54) (both at coexistence and both using the directly counted $g_{s,A}$ combinatorics of the self-avoiding polygons) with *no fit parameters*.

The right most plot of Fig. 9 also shows the integrated quantities of the density and pressure along the coexistence line for the $d=2$ Ising system. The values of $\rho_{\text{coex}}$ and $p_{\text{coex}}$ determined from calculations on the square lattice [211] (open circles), from Eqs. (21) and (18) (filled circles) and from Eqs. (21) and (54) (filled squares) are compared to the analytical solution of Onsager (solid line) [6]. One can still assume that the formation of clusters exhausts all the non-idealities and simply calculate the pressure and density from the self-avoiding polygon combinatorics and the finite cluster volume concentration, the equations

$$p_{\text{coex}} = T \sum_{s,A} g_{s,A} \exp\left[-\frac{s(a_s + 2p_{\text{coex}}l)}{T}\right] \text{ and } \rho_{\text{coex}} = \sum_{s,A} A g_{s,A} \exp\left[-\frac{s(a_s + 2p_{\text{coex}}l)}{T}\right]. \tag{58}$$

were solved iteratively using $a_s = 2$ and the directly counted $g_{s,A}$ [188]. As one might expect, at low temperatures, where the dilute phase is very dilute, the "ideal" expressions of Eqs. (21) and (18) work quite well. However as the temperature increases and more and more clusters appear in the dilute phase the "ideal" expressions fail and predict, as expected based on the cluster concentration predictions, pressure and density values that are higher than the Onsager solution [6]. The non-zero volume expressions of Eqs. (21) and (54) follow Onsager's solution and the Ising calculations more closely.

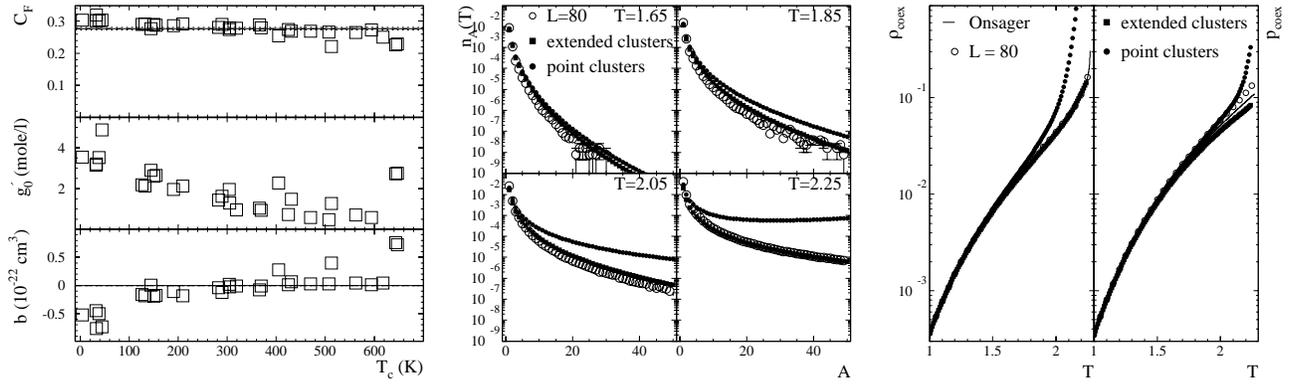

**FIGURE 9.** Left: The compressibility factor (top), normalization (middle) and excluded volume parameter (bottom) as a function of critical temperature for fluids from helium through water. Middle: Ising cluster yields from the $d=2$ square lattice (open circles) at four different temperatures compared to Eq. (18) (filled circles) and (54) (filled squares) (both at coexistence and both using the directly counted $g_{s,A}$ combinatorics of the self-avoiding polygons) with *no fit parameters* [200]. Right: the density $\rho_{\text{coex}}$ and pressure $p_{\text{coex}}$ at coexistence from the Onsanger solution (solid line), from $d=2$ Ising calculations on the square lattice [211] (open circles), from Eqs. (21) and (18) (filled circles) and from Eqs. (21) and (54) (filled squares).

This exercise shows that by leaving surface energy coefficient and critical temperature as a free parameters when fitting cluster concentrations, or by obtaining $T_c$ from other methods (e.g. the Onsager solution for the Ising model on a square lattice [6]), one accounts, for the most part, for the effects associated with the non-zero volume of the clusters.

*Super critical temperatures*

Finally, we note that Fisher's theory is valid only for $T \leq T_c$: temperatures greater than $T_c$ yield cluster surface free energies that are negative, and thus unphysical. The parametrization used in Fisher's theory is only one example of a more general form of the scaling assumption $n_A = A^{-\tau} f(X)$ and $X = A^{\sigma} \varepsilon^{\phi}$ and where $f(X)$ is some general scaling function which [38, 42, 46, 55, 160]:

- is valid on both sides of the critical point;
- for small $X$ ($T \sim T_c$ and small $A$) and $\varepsilon > 0$, $f(X)$ will vary as $\exp(-X)$ with $\sigma = 1/(\beta\delta) = 1/(\gamma + \beta) \sim 0.64$ for three dimensional Ising systems, 8/15 for two dimensional Ising systems or $\sim 0.45$ for three dimensional percolation systems and $\phi = 1$;
- for large $X$ ($T$ far from $T_c$ or large $A$) and $\varepsilon > 0$, $f(X)$ will vary as $\exp(-X)$ with $\sigma = (d-1)/d$ for all $d$ dimensional systems and with $\phi = 2\nu$.

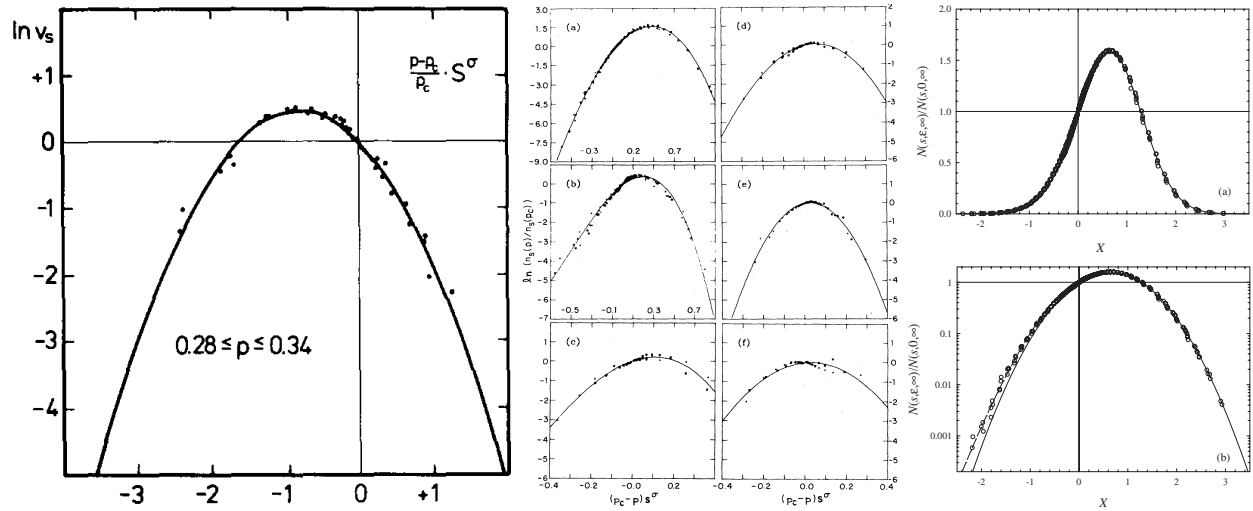

**FIGURE 10.** Left: the natural log of the scaled cluster yields $v_S = n_A(p)/A^{-\tau}$ as a function of the argument of the scaling function $X = A^\sigma(p - p_c)/p_c$ for bond building ($p$ is the bond building probability) $d = 3$ percolation on a square lattice with $10^6$ sites. Data from different $p$ follow the same curve as required by the scaling hypothesis. The parabola is the general form of $f(X)$ [42]. Middle: the natural log of the scaled cluster yields as a function of $p_c X$ (solid points) for (a) $d = 2$ to (f) $d = 7$ together with the least-squares fits (solid lines) [46]. Right: the scaled cluster yields plotted as a function of $X = \varepsilon A^\sigma$ for $|\varepsilon| \leq 4.5 \times 10^{-2}$ on a linear (a) and semi-logarithmic (b) scale; solid lines represents fits to a scaling function $f(X)$ [160].

Figure 10 shows the general form of the scaling function $f(X)$ for percolation systems [38, 42, 46, 160]. However, this more general scaling function $f(X)$ does not lend itself as easily to a physical interpretation as does the parameterization given by Fisher's theory and it is this physical interpretation which is important to the application of this method to the nuclear data.

## Summary

We have seen that Fisher's theory is a physical cluster theory. Fisher's main contribution was to introduce an accurate approximation for the entropic contribution to the cluster partition function. This lead to the development of a theory that shows both types of scaling: the singular behavior of quantities near that critical point and the scaling laws that relate exponents as well as the data collapse of cluster concentrations. Fisher's theory has an unphysical surface tension above the critical temperature, however below $T_c$ it serves as a good approximation that lends itself easily to a physical interpretation. Though Fisher's assumption about the mean surface of a cluster is crude (using a constant values for $a_0$ and $\sigma$ ignores the temperature dependence of the mean surface of a given cluster size) and it explicitly ignores the non-zero volume of the clusters (though implicitly the finite volume is almost all accounted for by the proper choice of $T_c$) it has successfully: described cluster production in percolating systems and Ising systems (see above); reproduced the compressibility factor at the critical point (see above); predicted (within a few percent) the compressibility factor of real fluids from the triple point to the critical temperature [30, 92]; and has been used to describe the nucleation rate of real fluids [32, 78].

## A BRIEF HISTORY OF NUCLEAR CLUSTER PRODUCTION

*ℐn the beginning there was neutron evaporation [4, 5], and the evaporation was good [12]...*

It was noted long ago that statistical methods could be applied to nuclear processes if the energies involved are large when compared to the lowest excitation energies of nuclei [4]. By doing this, Weisskopf was able to formulate expressions for the probability of neutron (or charged particle) emission from excited nuclei. Weisskopf based his work on evaporation from a body at low temperatures. In that regard, Weisskopf was working out the formulae to describe the evaporation of neutrons from a hot nucleus, i.e. he was describing a first order phase transition with a neutron leaving (or evaporating from) the condensed phase (the hot nucleus) and entering the dilute phase (a very low density neutron vapor).

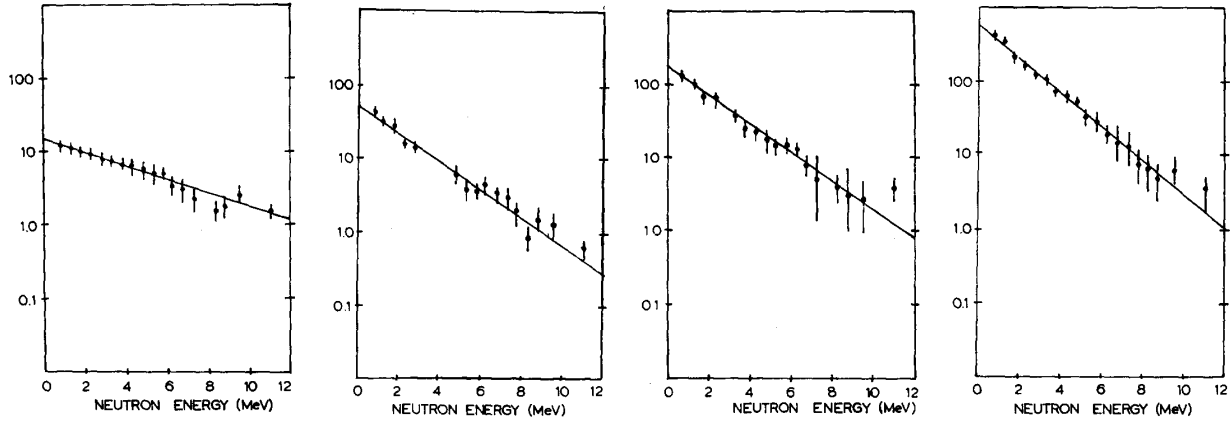

**FIGURE 11.** The scaled energy distributions of neutrons ($\frac{W_n(\mathscr{E})}{\sigma(E_0,\mathscr{E})\mathscr{E}^{1-i}}$ vs. $\mathscr{E}$) evaporated from (left to right): Al, Ni, Ag and Au nuclei after bombardment from 190 MeV protons [12]. The slopes of the lines give the inverse of the effective temperature of evaporation.

Following Bohr, Weisskopf divided processes initiated by nuclear collisions into two stages: the first was the formation of a compound nucleus and the second was the disintegration of the compound nucleus. Both stages could be treated independently. The energy of the compound nucleus is similar to the heat energy in a solid or liquid and the emission of particles from the compound nucleus is analogous to an evaporation process and Weisskopf derived a general statistical formula for the evaporation of particles from an excited compound nucleus (with the caveats of the finiteness of the nucleus and the fact that the evaporation of a particle takes away significant energy from the compound nucleus).

The probability per unit time of a nucleus $A_0$ with excitation energy $E^*$ emitting a neutron of mass $m_n$ with kinetic energy between $\mathscr{E}$ and $\mathscr{E}+d\mathscr{E}$ (where $d\mathscr{E}$ is much larger than the levels of $A_0$), thus transforming itself into nucleus $A_c$ with an excitation energy $E^* - E_0 - \mathscr{E}$ (where $E_0$ is the neutron binding energy of $A_0$) is

$$W_n(\mathscr{E})d\mathscr{E} = \sigma(E_0,\mathscr{E}) \frac{m_n \mathscr{E}}{\pi^2 \hbar^3} \exp\left\{-\frac{\mathscr{E} - T\left[\ln g + S_{A_0} - S_{A_c} - f(\mathscr{E})\right]}{T}\right\} d\mathscr{E} \tag{59}$$

where $\sigma(E_0,\mathscr{E})$ is the mean cross section for the collision of a neutron of kinetic energy $\mathscr{E}$ with nucleus $A_c$ of energy $E^* - E_0 - \mathscr{E}$ resulting in the production of nucleus $A_0$ of energy $E^*$; $g$ is the number of states for the spin of the particle under consideration; $S(E) = \ln \rho(E)$ corresponds to the entropy of a nucleus with and energy between $E$ and $E + dE$ (level density $\rho(E)$); $T$ is the temperature at which $E$ is the most probable energy of nucleus $A_c$; and $f(e)$ "contains all further terms of the development." The probability per unit time for the evaporation of particles of nucleon number $A$, charge $Z$ and mass $m_A$ from nucleus $A_0$ is

$$W_A(\mathscr{E})d\mathscr{E} = \pi R_0^2 \left(\mathscr{E} - e^2 \frac{Z_0 Z}{R_0}\right) \frac{m_A}{\pi^2 \hbar^3} \exp\left\{-\frac{\mathscr{E} + e^2 \frac{Z_0 Z}{r} - T\left[\ln g + S_{A_0} - S_{A_c} - f(\mathscr{E} - e^2 \frac{Z_0 Z}{R_0})\right]}{T}\right\} d\mathscr{E} \tag{60}$$

where $R_0$ is the radius of the compound nucleus and $Z_0$ is its charge. It is no surprise, given that Weisskopf had evaporation in mind, that equations (59) and (60) are similar to Fisher's estimate of the cluster partition function given in Eq. (14).

Multiplying the total probability of particle emission by $\hbar$ then gives the decay width: for neutrons:

$$\Gamma_n = \overline{\sigma} \frac{m_n}{\pi^2 \hbar^2} T^2 \exp\left(\ln g + S_{A_0} - S_{A_c}\right) \tag{61}$$

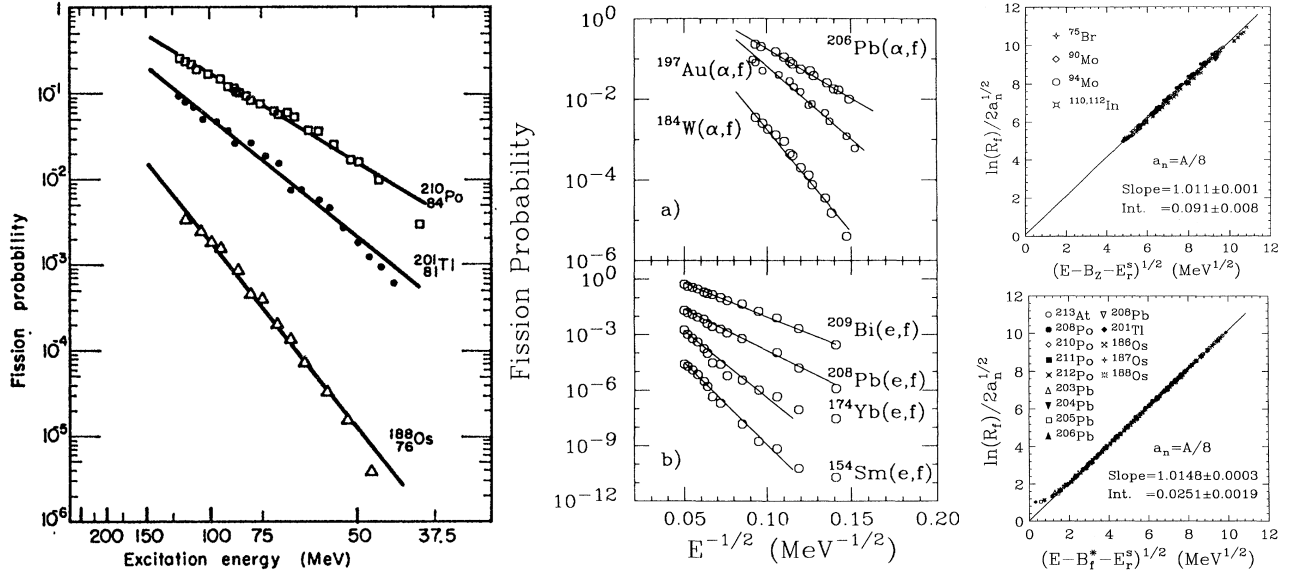

**FIGURE 12.** Left: The fission probability $P_f$ as a function of the inverse of the square root of the excitation energy $E^{*-1/2}$ for the reactions $^{206}_{82}\text{Pb}(^4\text{H},f)$, $^{197}_{79}\text{Au}(^4\text{H},f)$ and $^{184}_{74}\text{W}(^4\text{H},f)$ [21]. Middle: (a) The fission probability $P_f$ plotted as a function of $E^{*-1/2}$ for the $\alpha$ induced reactions $^{206}\text{Pb}(\alpha,f)$, $^{197}\text{Au}(\alpha,f)$ and $^{184}\text{W}(\alpha,f)$ and (b) the electron induced reactions $^{209}\text{Bi}(e,f)$, $^{208}\text{Pb}(e,f)$, $^{174}\text{Yb}(e,f)$ and $^{154}\text{Sm}(e,f)$ [85]. Right, top: the logarithm of the reduced mass-asymmetric fission rate $R_f$ divided by $2\sqrt{a_n}$ as a function of $E^{*-1/2}$ for the compound nuclei $^{75}\text{Br}$, $^{90}\text{Mo}$, $^{94}\text{Mo}$ and $^{110,112}\text{In}$ [98]. Right, bottom: the logarithm of the reduced mass-asymmetric fission rate $R_f$ divided by $2\sqrt{a_n}$ as a function of $E^{*-1/2}$ for the compound nuclei $^{186,187,188}\text{Os}$, $^{201}\text{Tl}$, $^{203,204,205,206,208}\text{Pb}$, $^{208,210,211,212}\text{Po}$ and $^{213}\text{At}$ [108].

(where $\overline{\sigma}$ is the mean value of $\sigma(E_0,\mathscr{E})f(\mathscr{E})$ averaged over the Maxwell distribution) and for charged particles

$$\Gamma_A = \sigma_0 \frac{m_p}{\pi^2\hbar^2} T^2 \exp\left(\ln g + S_{A_0} - S_{A_c}\right). \tag{62}$$

Thus Weisskopf developed a theory of nuclear evaporation, i.e. a theory of first order phase transition in finite, charged, asymmetric nuclear matter.

Experimental evidence of neutron evaporation appeared in the energy distributions of neutrons measured after various nuclei were bombarded with 190 MeV protons [12]. Equation (59) gives the probability of the evaporation of a single neutron from a single compound nucleus at a specific excitation energy. However, experimental measurements of neutron kinetic energy distributions were measured for neutrons that came from a cascade of successive evaporations from compound nuclei with a distribution of initial excitation energies. Thus to connect Eq. (59) with the experimental measurements the successive neutron (and proton) evaporation and distributions of initial excitation energies had to be taken into account which gives [12]

$$W_n(\mathscr{E})d\mathscr{E} \propto \sigma(E_0,\mathscr{E})\left(\frac{\mathscr{E}}{T}\right)^{i-1}\exp\left(-\frac{\mathscr{E}}{T}\right)\frac{1}{T}d\mathscr{E} \tag{63}$$

where $i$ is the generation of the evaporation. Figure 11 shows logarithmic plots of scaled neutron energy distributions ($\frac{W_n(\mathscr{E})}{\sigma(E_0,\mathscr{E})\mathscr{E}^{1-i}}$ vs. $\mathscr{E}$) follow a straight line whose slope is the inverse of the effective temperature of evaporation $T$. The plots in Fig. 11 are similar to the Arrhenius plots of nuclear cluster yields observed later [127], as such they present early evidence for thermal scaling in nuclear evaporation.

The thermal nature of cluster production in nuclear reactions was seen to extend all the way to fission [13, 21, 85, 98, 108]. The fission probability and fission cross section can be written as

$$P_f = \frac{\sigma_f}{\sigma_0} \text{ and } \sigma_f = \sigma_0 \frac{\Gamma_f}{\Gamma_T} \tag{64}$$

where

$$\Gamma_f = T \frac{\rho_s \left(E^* - B_f - E_r^s\right)}{2\pi\rho \left(E^* - E_r^{gs}\right)} \quad (65)$$

$\Gamma_f$ and $\Gamma_T$ are the fission and total decay widths, $B_f$ is the fission barrier, $\rho_s$ is the saddle-point level density, $\rho$ is the compound nucleus level density and $E_r^s$ and $E_r^{gs}$ are the saddle and ground-state rotational energies.

For large excitation energies ($E^* \gg B_f$, $E^* \gg E_r^s$ and $E^* \gg E_r^{gs}$) the fission width is

$$\Gamma_f = \frac{T\rho_s \left(E^* - B_f\right)}{2\pi\rho \left(E^*\right)} \simeq \frac{T\rho_s \left(E^*\right)}{2\pi\rho \left(E^*\right)} e^{-\frac{B_f}{T}} \simeq \frac{T}{2\pi} e^{-\frac{B_f}{T}}, \quad (66)$$

where the Boltzmann factor arise from the first-order term in the Taylor expansion of $\ln \rho_s \left(E^* - B_f\right)$. Then in the limit that the nucleus behaves as a Fermi gas with $T = \sqrt{E^*/a}$ the natural logarithm of the fission probability should go as

$$\ln P_f \simeq \ln \frac{\Gamma_f}{\Gamma_T} \simeq \mathscr{A} - \frac{\mathscr{B}}{\sqrt{E^*}} \quad (67)$$

where $\mathscr{A}$ and $\mathscr{B}$ are constants. Consequently, a plot of the natural logarithm of the fission probabilities versus $1/\sqrt{E^*}$ should be linear. This is just the case as is shown for several fission reactions in Fig. 12 [21, 85].

For lower excitation energies and nuclei with $B_f \gg B_n$ (the neutron evaporation barrier): $\Gamma_T \simeq \Gamma_n$. Then Eq. (64) can be rewritten as

$$\Gamma_n P_f \frac{2\pi\rho \left(E^* - E_r^{gs}\right)}{T} = \rho_s \left(E^* - B_f - E_r^s\right) \simeq e^{2\sqrt{a_f(E^* - B_f - E_r^s)}}, \quad (68)$$

where we have assumed a simplified form of the Fermi gas level density, and therefore,

$$\frac{1}{2\sqrt{a_n}} \ln \left[\Gamma_n P_f \frac{2\pi\rho \left(E^* - E_r^{gs}\right)}{T}\right] = \frac{\ln R_f}{2\sqrt{a_n}} = \sqrt{\frac{a_f}{a_n} \left(E^* - B_f - E_r^s\right)} \quad (69)$$

where $a_f$ and $a_n$ are the level density parameters associated with the fission saddle point and the ground state and $R_f$ is the reduced mass-asymmetric fission rate. The neutron width can be approximated as

$$\Gamma_n \simeq \frac{2m_n R^2 g}{\hbar^2} T_n^2 \frac{\rho \left(E^* - B_n - E_r^{gs}\right)}{2\pi\rho \left(E^* - E_r^{gs}\right)} \quad (70)$$

where $B_n$ is the last neutron binding energy, $T_n$ is the temperature after neutron evaporation and $R$ is the radius of the compound nucleus.

For fission excitation functions in the Pb region, strong shell effects make the approximation

$$\rho \left(E - B_n - E_r^{gs}\right) \propto e^{2\sqrt{a_n\left(E^* - B_n - E_r^{gs}\right)}} \quad (71)$$

a very poor one. However, for excitation energies higher than 15–20 MeV, the level density assumes its asymptotic form [33]:

$$\rho \left(E - B_n - E_r^{gs}\right) \propto e^{2\sqrt{a_n\left(E^* - B_n - E_r^{gs} - \Delta_{shell}\right)}}, \quad (72)$$

where $\Delta_{shell}$ is the ground-state shell effect of the daughter nucleus after neutron evaporation. Assuming this asymptotic expression for the level density after neutron evaporation, the fission excitation functions can be fit with Eq. (64) using $\Delta_{shell}$ as a free parameter in the expression for $\Gamma_T \simeq \Gamma_n$ [108]. Thus, a plot of the left hand sige of Eq. (69) (which can be constructed from measured fission cross sections and known non fission channels (mostly neutron evaporation)) versus $\sqrt{E^* - B_n - E_r^s}$ should be linear (actually a 45° line for $a_f = a_n$). That this is so can be seen in Fig. 12 [98, 108], where a large number of fission excitation functions scale exactly to the same straight line illustrating the thermal nature of cluster production in nuclear fission reactions.

If the analogous behavior of evaporation from excited nuclei and evaporation of classical fluids holds, then one expects that as the temperature increases the first order phase transition (evaporation) becomes a continuous phase transition at a critical temperature $T_c$ above which there is a smooth cross over from the liquid-like phase of ordinary nuclear matter encountered at low excitation energies to a gaseous phase where the average interparticle distance is much larger that the range of the interparticle interaction. Thus, when inclusive (i.e. the average cluster yield for a given cluster size $A$ was generated by averaging over all excitation energies) cluster yields from the reaction of $80 \leq E_{\text{beam}} \leq 350$ GeV protons incident on krypton and xenon nuclei exhibited a power law (as expected for $n_A(T_c)$ in Eq. (18)) with an exponent $\tau$ between 2 and 3 (as expected for $d = 3$ systems [42]) it seemed possible that the critical temperature had been reached [50, 51, 53, 54].

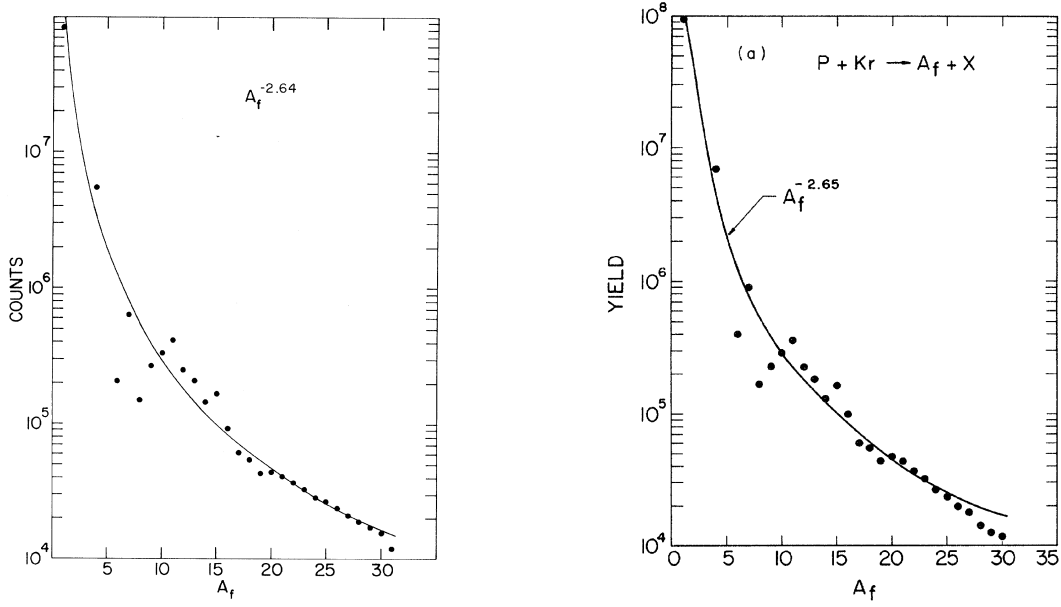

**FIGURE 13.** The inclusive cluster yields $n_A$ as a function of cluster size $A$ from the reaction of $80 \leq E_{\text{beam}} \leq 350$ GeV protons incident on xenon (left [50]) and krypton (right [54]) nuclei. Solid lines show a fit to the data with a power law.

**TABLE 1.** Values for the parameters in the fits to the isotopic yields. These values were calculated by fixing $a_V = 14.1$ MeV and leaving the remaining parameters free [51, 54].

| Parameter | $a_V$ (MeV) | $a'_s$ (MeV) | $a_C$ (MeV) | $a_a$ (MeV) | $a_p$ (MeV) | $\mu_P$ (MeV) | $\mu_N$ (MeV) | $T$ MeV | $\tau$ |
|---|---|---|---|---|---|---|---|---|---|
| **Nominal value** | 14.10 | 13.00 | 0.60 | 19.00 | 33.50 | | | | 2.21 |
| **p+Kr** | 14.10 | 5.53 | 0.49 | 22.66 | 5.92 | $-11.32$ | $-7.59$ | 3.28 | 2.64 |
| **p+Xe** | 14.10 | 6.61 | 0.40 | 23.30 | 5.28 | $-11.01$ | $-7.62$ | 3.24 | 2.65 |

An analysis of the isotopic cluster yields provided further evidence that the clusters arising from the p+Xe and p+Kr reactions were thermal in nature and perhaps critical. The measured inclusive yields were fit to a version of Fisher's theory modified to account for the nuclear aspects of the fluid in question. Specifically the yields $Y(A,Z)$ of a cluster with $A$ nucleons, $N$ neutrons and $Z$ protons were fit with the following parameterization [51, 54]

$$Y(A,Z) = CA^{-\tau} \exp\left[\frac{a_V A - a'_s A^{2/3} - a_C \frac{Z^2}{A^{1/3}} - a_a \frac{(A-2Z)^2}{A} - \delta + \mu_N N + \mu_Z Z}{T} + N\ln\frac{N}{A} + Z\ln\frac{Z}{A}\right] \quad (73)$$

with $\delta = a_p A^{-3/4}$ for odd-odd nuclei, $\delta = 0$ for odd-even nuclei and $\delta = -a_p A^{-3/4}$ for even-even nuclei. Here Fisher's theory has been modified to use Weizäcker's semiempirical mass formula (the first five terms in the Boltzmann factor) and a chemical potential for neutrons and protons (the last two terms in the Boltzmann factor). The final terms in the exponential in Eq. (73) take into account the entropy of mixing protons and neutrons. Figure 14 shows the results for the 59 different isotopes from the p+Kr reactions and 62 different isotopes from the p+Xe reaction fit to Eq. (73) with the free parameters and results given in Table 1.

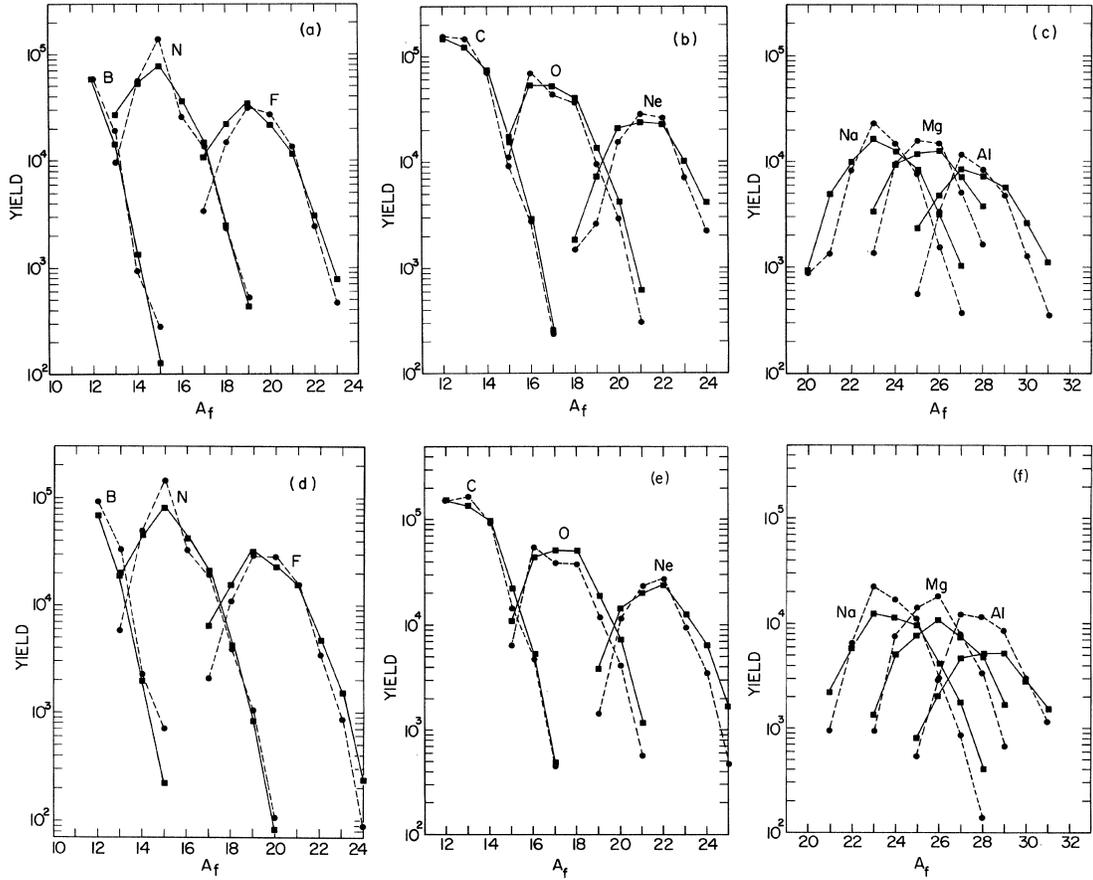

**FIGURE 14.** Top (bottom): the isotopic inclusive cluster yields $Y(A,Z)$ as a function of cluster size $A$ from the reaction of $80 \leq E_{\text{beam}} \leq 350$ GeV protons incident on xenon (krypton) nuclei [51, 54]. Circles represent data, while squares are the fit according to Eq. (73). The dashed and solid lines are drawn to guide the eye.

In a similar analysis [70], the inclusive cluster yields ($3 \leq Z \leq 14$) from the reaction of p+Xe was fit to Fisher's theory in a more generic form

$$n_A(T) = g'_0 A^{-\tau} X^{A^{\sigma}} Y^A \qquad (74)$$

where the surface free energy contribution is found in $X = \exp(-a'_s \varepsilon / T)$ and the chemical potential contribution is found in $Y = \exp(\Delta\mu/T)$. The energy of the incident proton was varied from $E_{\text{beam}} = 1$ GeV to $E_{\text{beam}} = 18$ GeV. At each beam energy the inclusive cluster yields were fit to Eq. (74) with the normalization and $X$ and $Y$ left as free parameters and the exponents $\sigma$ and $\tau$ were fixed to their $d = 3$ Ising values. Figure 15 suggests that at low beam energies the system is super saturated ($Y \geq 1$ indicates that $\Delta\mu \geq 0$) and there is a sizeable surface free energy cost in cluster formation. As the beam energy is increased, the system moves towards coexistence ($Y \to 1$) and the surface free energy vanishes ($X \to 1$). For beam energies above 12 GeV $X = Y = 1$ and the cluster yields are well fit by a power law.

While the results of the analysis of the above two experiments are not completely clear, it is clear that the clusters created in these reactions are well described as a thermal phenomenon. Regardless of whether the critical point was reached, the inclusive cluster yields for three different reactions over a wide range in excitation energies were well fit by Fisher's theory. Using Fisher's theory to describe clusters emitted from highly excited nuclei was, in hindsight, a natural extension of the theoretical work of Weisskopf [4] and the experimental work on neutron evaporation [12]. Specifically the extension of Weisskopf's particle evaporation probabilities to include Fisher's estimates of the entropic cost of cluster formation is much the same as the actual development of physical cluster theories [190].

The above experimental results stimulated much theoretical interest in the possibility of critical phenomena in nuclear matter [52, 57, 58, 59, 63, 64, 65, 71, 72, 81, 84, 87, 88, 90, 91, 93, 96, 97, 99, 104, 105, 109, 110, 118, 121,

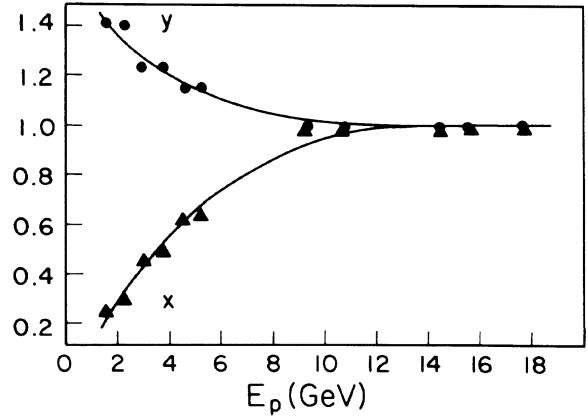

**FIGURE 15.** The parameters $X$ (related to the surface free energy) and $Y$ (related to the chemical potential) obtained from a fit of inclusive cluster yields ($3 \leq Z \leq 14$) from the reaction of p+Xe ($1 \leq E_{\text{beam}} \leq 20$ GeV) to Eq. (74) [70].

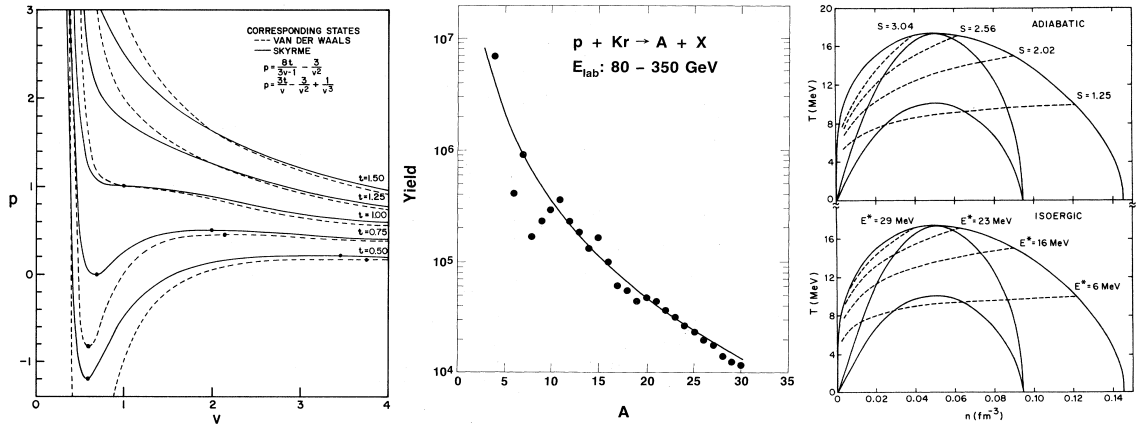

**FIGURE 16.** Left: the pressure as a function of volume; the dashed curves show the van der Waals fluid and the solid curves show the results for a system interacting through a Skryme force [52]. Middle: The nuclear cluster yields of reference [50] fit with the theory of refrence [58]. Right: the outer solid curve is the temperature-density coexistence curve, the middle solid curve is the isothermal spinodal and the inner solid curve is the isentropic spinodal; dashed curves show trajectories for constant entropy expansion (top) and constant energy expansion (bottom) [65].

123, 133, 143, 150, 153, 163, 181, 186]. These efforts can be separated into two different categories: analytical/semi-analytical theories [52, 57, 58, 63, 64, 65, 105] and computational models: both on a lattice [59, 71, 72, 81, 84, 87, 88, 93, 96, 104, 110, 121, 123, 133, 143, 153, 163, 181] and off [90, 91, 97, 99, 109, 111, 118, 122, 124, 136, 141, 150, 175, 180, 182, 185, 186, 214].

The analytical/semi-analytical theories employed various methods (e.g. particles interacting through a Skyrme force, finite temperature Hartree-Fock theory and another nuclear extension of Fisher's theory) to determine the critical point of bulk (i.e. infinite, uncharged and symmetric) nuclear matter and the liquid-vapor phase boundary. This lead to estimates of the critical temperature in the range of 12.6 MeV to 28.9 MeV depending on the theoretical techniques employed. Once estimates were made for bulk nuclear matter, the effects of a finite number of nucleons and a fluid made up of two components (one which carries an electric charge) were studied. Those effects generally lead to a lower critical temperature with estimates between 8.1 MeV and 20.5 MeV.

Computational models on the lattice attempted to study the process of nuclear cluster formation from "the bottom up" by modeling in a simple way the short range interaction of the nucleons. This was done both geometrically with perolation models [59, 71, 72, 81, 84, 87, 88, 123] and thermally with lattice gas (Ising) models [93, 96, 110, 121, 133, 143, 153, 163, 181].

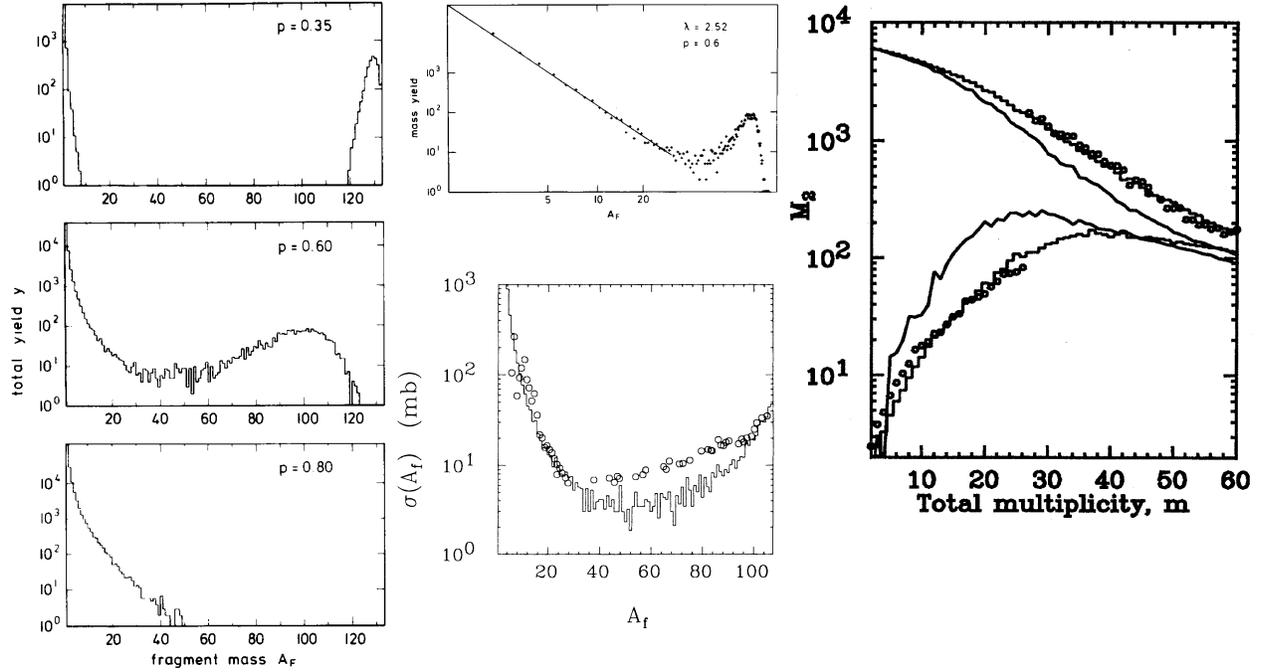

**FIGURE 17.** Results from from percolation models. Left: the dependence of the cluster yields as a function of the bond breaking probability (top: $q = 0.35$, middle: $q = 0.60$, bottom: $q = 0.80$) for the reaction p+Xe [59]. Middle top: a doubly logarithmic plot of the mass yield for $q = 0.6$ for the reaction p+Xe, the straight line shows the power law $A^{-2.52}$ [59]. Middle bottom: a comparison of the data from ref. [62] (open circles) and the percolation model [72] (solid line). Right: the second moment of the cluster distribution as a function of the cluster multiplicity for the reaction of 1 AGeV Au+C: histogram: percolation [104]; circles: data [89].

The percolation model describes the excited nucleus in question as a lattice with nucleons at every vertex. The distance between each vertex, or the lattice spacing, depends on the density of normal nuclei, $\rho_0$ and is approximately $\rho_o^{1/3} \sim 1.8$fm. Bonds between the nucleons are broken with a probability $q(E^*)$ that depends on the excitation energy $E^*$ of the system: the greater the $E^*$ the higher the value of $q$ [59, 72, 87]. This mapping of excitation energy onto bond breaking probability is similar to the mapping of a ferromagnetic Potts model onto a corresponding percolation model [25, 29, 45, 87]. Thus the percolation model becomes an Ising model and the success in describing a thermal phenomenon such as cluster production from an excited nucleus is to be expected. Figure 17 shows that the percolation model was able to reproduce many of the observations of the experimental measurement: the cluster yields were described by a power law at some value of $q(E^*)$ and matched experimental measurements. The model also gave indications of what would be expected for exclusive cluster yields, i.e. cluster yields that could be separated based on some measure of the excitation energy of the reaction. The percolation model was compared with data from many other reactions and studied the influence of the shape of the lattice boundary (e.g. spherical [81] and toroidal [84]) and has successfully described the clusters arising from excited gold like nuclei [104, 177].

A complete (every cluster measured), exclusive data set of the clusters from the reaction 900 AMev Au+emulsion [60] was compared to the clusters from bond percolation on the simple cubic lattice with 216 sites [67, 71]. The moments from each cluster distribution were calculated as

$$M_k(T) = \sum_{A=1}^{A < A_{\max}} A^k n_A(T). \tag{75}$$

where $A_{\max}$ is the size of the largest cluster in a given event or lattice realization. In the case of the percolation moments, $T$ was replaced with the bond probability $q$ according to common practice [42]. In the case of the nuclear moments, $T$ was replaced by the total cluster multiplicty $m$ (more specifically the total charged particle multiplicity) and the nucleon number $A$ of the cluster was replaces with the charge of the cluster $Z$. The moments of the cluster distribution were used to determine the value of $\tau$ by plotting the third moment $M_3$ as a function of the second moment

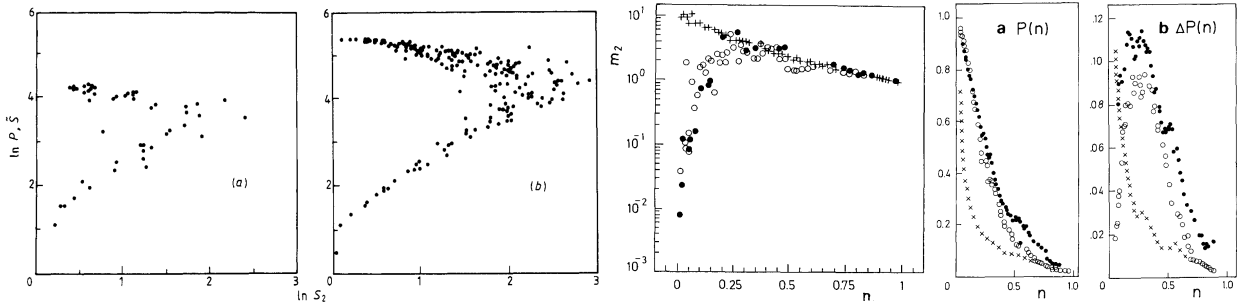

**FIGURE 18.** From left to right: a plot of the natural log of the third moment of the cluster distribution as a function of the natural log of the second moment for nuclear clusters [60, 67]; a plot of the natural log of the third moment of the cluster distribution as a function of the natural log of the second moment for bond percolation on the simple cubic lattice [67]; the second moment of the cluster distribution as a function of the total cluster multiplicity for nuclear data (full circles), percolation on the simple cubic lattice (open circles) and percolation on a line (crosses) [71]; the size of the largest cluster as a function of the total cluster multiplicity (see previous plot for symbol definition); and the fluctuations in the size of the largest cluster as a function of total cluster multilpicity [71]. See text for discussion.

$M_2$ and recalling (see above discussion on scaling from Fisher's theory or reference [42]) that for

$$M_k \propto |\varepsilon|^{\frac{\tau-1-k}{\sigma}} \tag{76}$$

so that $M_3 \propto M_2^{\frac{\tau-4}{\tau-3}}$. Thus the slope of a plot of $\ln M_3$ as a function of $\ln M_2$ is related to the exponent $\tau$ and for both percolation clusters and nuclear clusters this analysis yielded $\tau = 2.2 \pm 0.2$ [67]. On a more qualitative level, a plot of the natural log of the size of the largest cluster $A_{\max}$ (or the charge of the largest nuclear cluster $Z_{\max}$) as a function of $\ln M_2$ shows two branches. This is also shown in Fig. 18 where the upper branch can be thought of as the condensed phase (all particles in a single large cluster) and the lower branch can be thought of as the dilute phase (all particles in small clusters). Finally, comparisons of the second moment, largest cluster size and fluctuations in the largest cluster size all as a function of total cluster multiplicity shown in Fig. 18 show a similarity between the nuclear data (full circles) and $d = 3$ percolation (open circles) but not with $d = 1$ percolation (crosses) a system without a phase transition [42]. The conclusion of this analysis was that nuclear clusters are produced from a system that behaves as finite systems which have a phase transition in the infinite limit. Behavior of this sort for the second moment of the cluster distribution and size of the largest cluster has been observed in many experiments [86, 144, 151, 162, 183, 189, 192, 193, 212].

In reactions of 600 AMeV Au on various targets (C, Al, Cu and Pb) a high degree of universal scaling behavior was observed [80, 82, 86]. Figure 19 shows the behavior of the mean charge of the largest cluster $\langle Z_{\max} \rangle$, the value of $\tau$ of the cluster distribution, the mean number of intermediate mass clusters (where an IMC is defined as clusters with charge $3 \leq Z \leq 30$) $\langle M_{\text{IMC}} \rangle$ and the mean longitudinal velocity of a cluster $\langle \beta_\| \rangle$ and the ratio of the clusters' root mean square deviations of the transverse and longitudinal velocity $\text{rms}(\beta_\perp)/\text{rms}(\beta_\|)$ all as a function of the violence of collision (the more violent the collision, the higher the temperature). The universal scaling behavior associated with measures of the cluster yields indicated that the cluster yields did not depend on the target but on the energy deposited by the collision and is a necessary—though not sufficient—condition for chemical equilibrium being established. The universal scaling behavior of $\langle \beta_\| \rangle$ and $\text{rms}(\beta_\perp)/\text{rms}(\beta_\|)$ are compatible with the assumption of a kinetic equilibrium being accomplished prior to the decay of the primary spectator [82]. The universal scaling behavior shown in Fig. 19 supports the idea of equilibrium that Weisskopf (following Bohr) had in mind in his neutron evaporation work [4].

An analysis of the clusters with $Z > 5$ produced in the reactions 60 AMeV Au+Al, V and Cu showed that the natural logarithm of the branching ratios for binary, ternary, quaternary and quinary decay depended linearly on $E^{*-1/2}$ strongly suggesting the clusters were produced statistically [85]. These results were the natural extension of the analysis of Weisskopf [4]. This can be seen by assuming that $B_2, B_3, B_4, \ldots, B_n$ are the average "barriers" associated with binary, ternary, quaternary and quinary decays (i.e. a reaction at a given value of $E^*$ results in one, two, three, or four clusters and the residual nucleus). The decay probability $P_n$ for each channel is proportional to the level density of the system $\rho(E^*)$ as

$$P_n(E^*) \propto \rho(E^* - B_n) \tag{77}$$

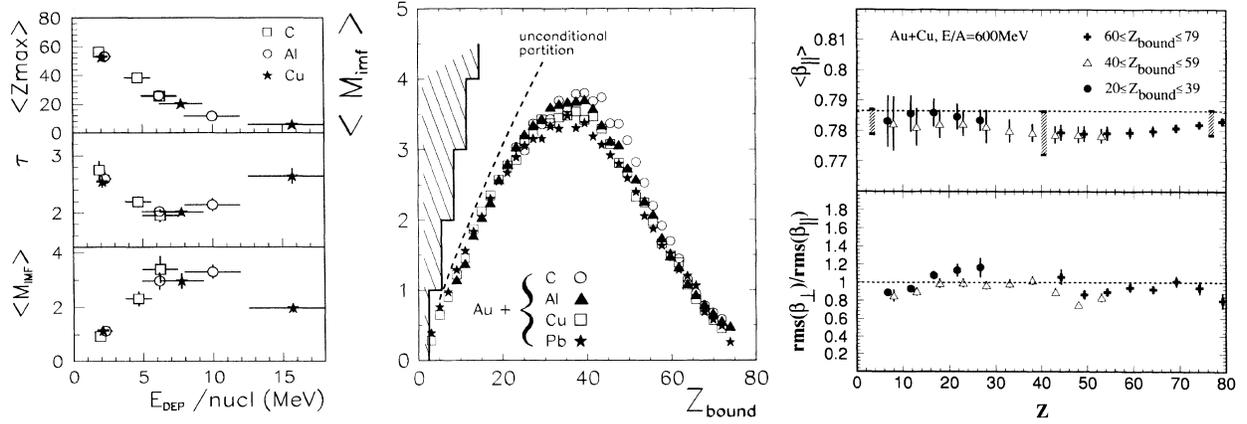

**FIGURE 19.** Left: From top to bottom: the mean charge of the largest cluster $Z_{max}$, the value of $\tau$ of the cluster distribution, the mean number of intermediate mass clusters (where an IMC is defined as clusters with charge $3 \leq Z \leq 30$) $\langle M_{IMC} \rangle$ plotted as a function of the estimate of the energy deposited into the excited nucleus [80]. Middle: $\langle M_{IMC} \rangle$ plotted as a function of the the summed charge for all clusters with $Z \geq 2$ $Z_{bound}$ [82]. Right: The mean longitudinal velocity of a cluster $\langle \beta_\parallel \rangle$ (top) and the ratio of the clusters' root mean square deviations of the transverse and longitudinal velocity $\mathrm{rms}(\beta_\perp)/\mathrm{rms}(\beta_\parallel)$ (bottom) plotted as a function of the charge of the cluster Z, different symbols show different bins in $Z_{bound}$ [82].

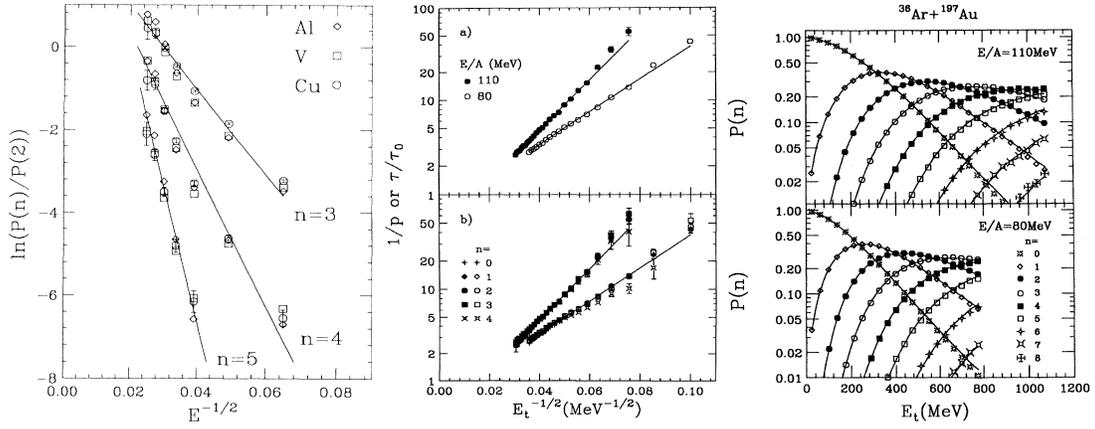

**FIGURE 20.** Left: The natural logarith of the threefold, fourfold and fivefold probabilities normalized to the twofold probability (symbols) as a function of $E^{*-1/2}$, lines are best fits to the data [85]. Middle: (a) The reciprocal of the binary decay probability $1/P_2$ or (b) the ratio $t/t_2$ as a function of the square root of the transverse energy $E_t^{-1/2}$, solid lines are fits to the data in the upper panel only [95]. Right: The experimental (symbols) and the calculated (solid) probability to emit $n$ intermediate mass clusters (IMCs) as a fucntion of the transverse energy $E_t$. For $n = 0 - 8$, $P_n^m$ ($P(n)$ in the figure) is calculated assuming a binomial distribution with the values of $P_2$ from the linear fits shown in the plots immediately to the left and the corresponding values of $m$ from Eq. (84) [95].

Using a Fermi gas level density with a constant level density parameter $a$ and in the limit that $E^* \gg B_n$ gives

$$P_n(E^*) \propto \exp\left(-\frac{B_n}{T}\right). \tag{78}$$

Figure 20 shows that the ratio of $n$-fold events to binary events

$$\ln\left(\frac{P_n}{P_2}\right) \propto -\sqrt{\frac{a}{E^*}}(B_n - B_2). \tag{79}$$

is linear in $E^{*-1/2}$ which is proportional to $T$. The linearity of $\ln(P_n/P_2)$ as a function of some measure of the temperature is called thermal scaling and is an indication that these clusters were created thermally.

A related analysis of cluster multiplicity distributions for the reactions 80 and 110 AMeV Ar+Au reactions exhibited binomial behavior at all excitation energies [95]. That is, a single binary event probability $P_2$ could be extracted with a thermal dependence indicating that cluster production is reducible to a combination of nearly independent emission processes. Once again this arises naturally from Weisskopf's work on nuclear evaporation [4]. The partial decay width for a given binary channel is approximately

$$\Gamma_2 \simeq \hbar \omega_2 \exp\left(-\frac{B_2}{T}\right) \quad (80)$$

where $\omega_2$ is a frequency characteristic of the binary decay channel. In fission, $\omega_2$ is the collective frequency of assault on the barrier and $B$ is the fission barrier. The binary decay probability is related to the partial decay width

$$P_2 \simeq \frac{\Gamma_2}{\hbar \omega_2}. \quad (81)$$

The channel period is $t_2 = 1/\omega_2$ and the corresponding decay time is

$$t \simeq t_2 \exp\left(\frac{B_2}{T}\right). \quad (82)$$

For nuclei with small $E^*$ (e.g. compound nuclei) the total decay width is the sum of the widths over all channels. For nuclei with larger $E^*$ only the decay width of the binary channel need be considered, while the abundant light particle decay can be treated as a background that may modify $T$ and $B_2$.

If we assume that the excited nucleus has the opportunity to try $m$ times to emit a cluster with constant $P_2$ probability of success, then the probability $P_n^m$ of emitting exactly $n$ decay products ($n-1$ clusters and the residual nucleus) is given by the binomial distribution

$$P_n^m = \frac{m!}{n!(m-n)!} (P_2)^2 (1-P_2)^{m-n}. \quad (83)$$

The average multiplicity and variance are then

$$\langle n \rangle = mP_2 \text{ and } \sigma_n^2 = \langle n \rangle (1-P_2) \quad (84)$$

thus one can extract the values of $P_2$ and $m$ directly from experimental measurements of the mean multiplicity and its variance at any excitation energy. This is shown in Fig. 20 for data from the reaction of $^{36}$Ar+$^{197}$Au. One can also extract $P_2$ "differentially" from the ratios of $\frac{P_n^m}{P_{n+1}^m}$ from

$$\frac{1}{P_2} = \frac{t}{t_2} = \frac{P_n^m}{P_{n+1}^m} \frac{m-n}{n+1} + 1. \quad (85)$$

These results are also shown in Fig. 20 for data from the reaction of $^{36}$Ar+$^{197}$Au. Both the method of measuring the mean multiplicity and variance and the differential method show a linear relation to the square root of the transverse energy $E_t$. $E_t$ is defined as $\sum e_i \sin^2 \theta_i$, where $e_i$ is the kinetic energy of the $i^{\text{th}}$ particle detected in an event and and $\theta_i$ is the angle between the particle and beam direction [95] and is proportional to the excitation energy $E^*$ which is proportional to $T^2$, thus $T \propto \sqrt{E_t}$. The thermal scaling of $\ln(1/P_2)$ (or $\ln(t/t_2)$) is an indication that these clusters were created thermally.

Figure 20 also shows a comparison between the experimental excitation functions and those calculated using the vales of $P_2$ from the linear fits in Fig. 20 and the associated values of $m$ from Eq. (84). The quantitative agreement between calculations and the experimental data confirm the binomality of the process which created these clusters and demonstrates that the probability of producing $n-1$ clusters, $P_n$ is reducible to the probability of producing one cluster, $P_2$. This type of reducibility is a strong indication that the clusters were created independently of each other.

The above signatures have come to be called reducibility (the probability of the production of $n$ clusters is reducible to the probability of producing a single cluster) and thermal scaling (the natural logarithm of the cluster yields is proportional to an inverse of some measure of the temperature). The presence of these signatures has been amply verified in nuclear reactions [85, 95, 101, 106, 115, 127, 135, 149, 154, 155, 159, 164, 173] and has been shown to be present in percolation [159], Ising [161, 180, 198] and classical molecular dynamics models [214] as well as inherent in Fisher's theory [159, 198].

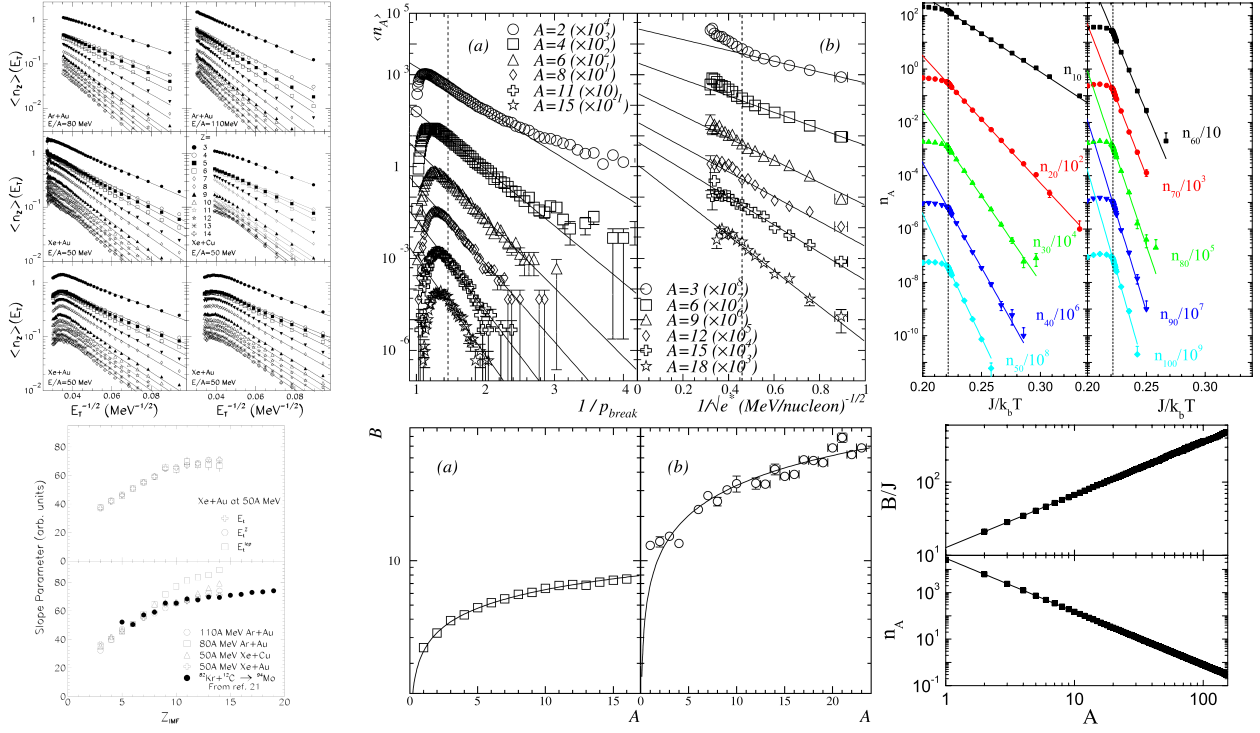

**FIGURE 21.** Left: Top: Arrhenius plots of the cluster charge yield of various reactions (see legend) distributions as a function of the square root of the inverse of the transverse energy $E_t$, solid lines show a fit to $Y(Z, E_t) = B_0 \exp\left(-B_Z/E_t^{1/2}\right)$. Bottom: The behavior of the extracted barrier $B_Z$ as a function of the charge $Z$ of the cluster [135]. Middle: Top, left: Arrhenius plots of the cluster yield distributions from bond percolation on a simple cubic lattice as a function of the bond breaking probability $q$, solid lines show a fit to $Y(A, q) = B_0 \exp(-B_A/q)$. Bottom, left: The behavior of the extracted barrier $B_A$ as a function of the cluster number, solid lines show a fit to $B_A = a'_s A^\sigma$, both $a'_s$ and $\sigma$ were in agreement with the expected values for $d = 3$ percolation. Top, right: Arrhenius plots of the cluster yield distributions from the reaction 1 AGeV Au+C as a function of the square root of the inverse of the excitation energy $E^*$, solid lines show a fit to $Y(A, E^*) = B_0 \exp\left(-B_A/E^{*1/2}\right)$. Bottom, right: The behavior of the extracted barrier $B_A$ as a function of the cluster number, solid lines show a fit to $B_A = a'_s A^\sigma$, the value of $\sigma$ was in agreement with the expected values for $d = 3$ Ising value and the value of $a'_s$ was roughly half of the expected value for nuclear matter [159]. Right: Top: Arrhenius plots of the cluster yield distributions from a simple cubic Ising lattice as a function of the temperature $T$, solid lines show a fit to $Y(A, T) = B_0 \exp(-B_A/T)$. Right, bottom: The behavior of the extracted barrier $B_A$ as a function of the cluster number, solid lines show a fit to $B_A = a'_s A^\sigma$, both $a'_s$ and $\sigma$ were in agreement with the expected values for $d = 3$ Ising systems (also shown is the power law in $n_A(T_c)$ at the critical point) [198].

Fisher's theory shows thermal scaling quite clearly. Begining from the cluster number concentration as given in Eq. (18) and working at coexistence ($\Delta\mu = 0$) we can immediately write

$$n_A(T) = g'_0 A^{-\tau} \exp\left(\frac{a'_s A^\sigma}{T_c}\right) \exp\left(-\frac{a'_s A^\sigma}{T}\right) = B_0 \exp\left(-\frac{B_A}{T}\right) \tag{86}$$

where $B_0$ contains all the temperature independent terms and $B_A$ is the barrier associated with the production of a cluster of $A$ constituents. Equation (86) shows that the barrier should increase with increasing cluster number: $B_A = a'_s A^\sigma$. This behavior was observed in a wide variety of heavy ion collisions over a broad range of energies when the natural logarithm of the yield of clusters of a given charge were fit to $Y(Z, E_t) = B_0 \exp\left(-B_Z/E_t^{1/2}\right)$ where $E_t$ is the transverse energy [135]. The left column of Fig. 21 shows the fits to $Y(Z, E_t)$ and the behavior of the extracted barrier $B_Z$ as a function of the charge of the cluster in question. The middle column of Fig. 21 also shows the barriers $B_A$ determined from the cluster yields as a function of bond breaking probability $q$ for bond percolation on the simple cubic lattice ($Y(A, q) = B_0 \exp(-B_A/q)$) and from the cluster yields as a function of the square root of the excitation energy $E^*$ ($Y(A, E^*) = B_0 \exp\left(-B_A/E^{*1/2}\right)$) for the reaction 1 AGeV Au+C. For both percolation and the nuclear reaction the barrier $B_A$ was observed to vary as $a'_s A^\sigma$ with $a'_s$ and $\sigma$ equal to their percolation values for the percolation clusters and $\sigma$ equal to its $d=3$ Ising value for the nuclear clusters and $a'_s$ roughly half its expected value for nuclear matter [159]. The right column of Fig. 21 shows the barriers $B_A$ determined from the cluster yields from a simple cubic Ising lattice as a function of temperature ($Y(A,T) = B_0 \exp(-B_A/T)$) [198]. Again the value of the barrier $B_A$ went as $a'_s A^\sigma$ with both $a'_s$ and $\sigma$ close to their expected values. In all cases clusters of a wide range in size (as measured by $Z$ or $A$) and over a wide range in "temperature" (as measured by $E_t$, $E^*$, $q$ or $T$) showed thermal scaling.

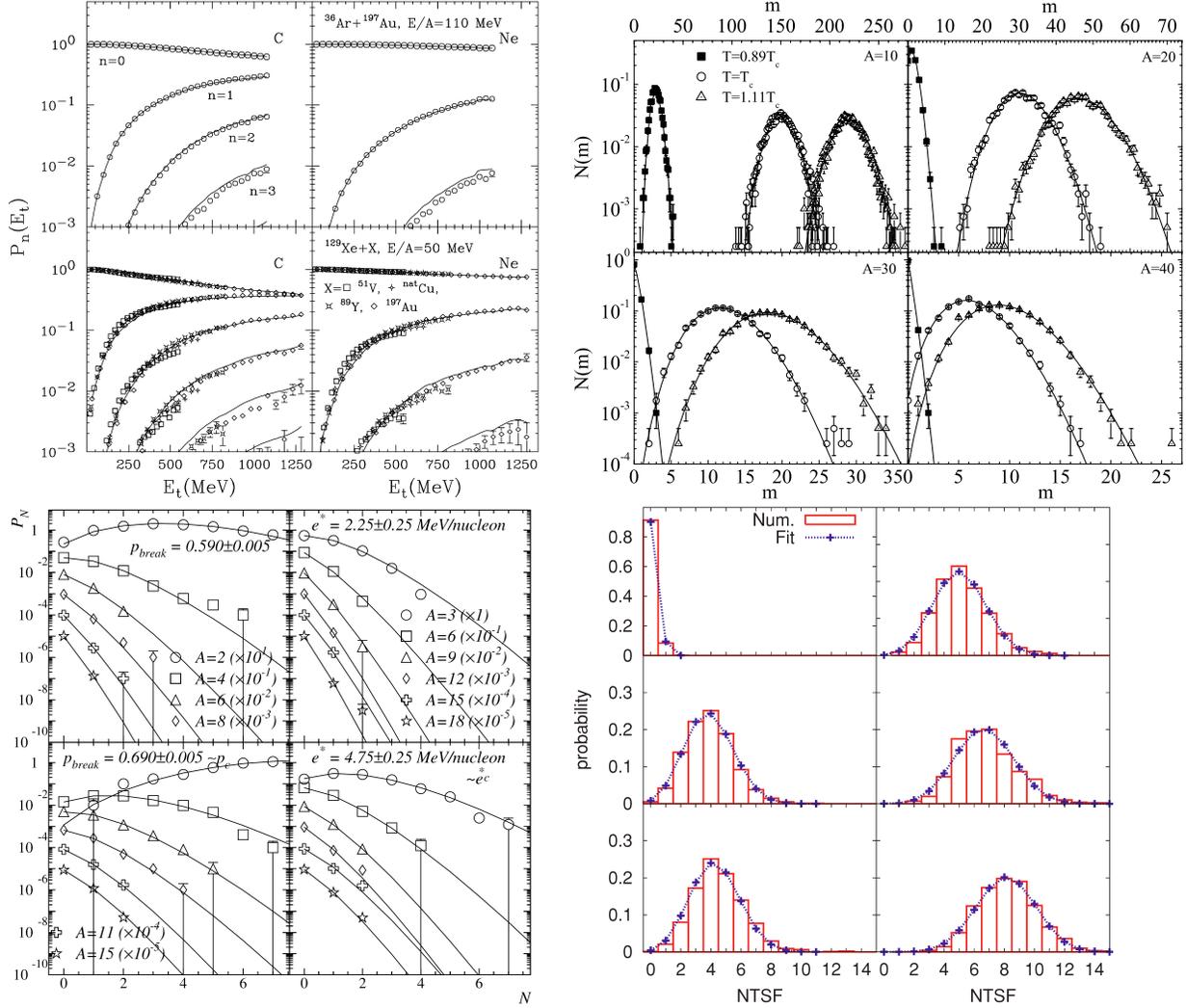

**FIGURE 22.** Top, left: the excitation functions $P_n$ for carbon (left column) and neon (right column) emission from reactions $^{36}$Ar+$^{197}$Au at 100 AMeV (top panels) and $^{129}$Xe+$^{51}$Vi, $^{nat}$Cu, $^{89}$Y, $^{197}$Au (bottom panels); the lines are Poissonian fits to the gold target data [135]. Bottom, left: the multiplicity distributions $P_N$ for clusters of size $A$ as a fucntion of $N$ in bins of bond breaking probability $p_{break}$ and excitation energy $E^*$ for percolation (left) and the reaction 1 AGeV Au+C (right); lines are Poissonian distributions calculated with the measure $\langle N_A \rangle$ [159]. Top, right: the probability distributions of obtaining $m$ clusters of size $A$ at the three temperatures indicated; lines are fits to a Poissonian distribution (Eq. (87) with the means given by the data [198]. Bottom, right: the probability distributions (histograms) and binomial fit (Eq. (83) dotted line) for the production of NTSF number of clusters at increasing energies from low (upper left) to highest (lower right) for classical molecular dynamics calculations [214].

Reducibility: Poissonian for the case of infinite systems

$$P_n = \frac{\langle n \rangle}{n!} e^{-\langle n \rangle}. \tag{87}$$

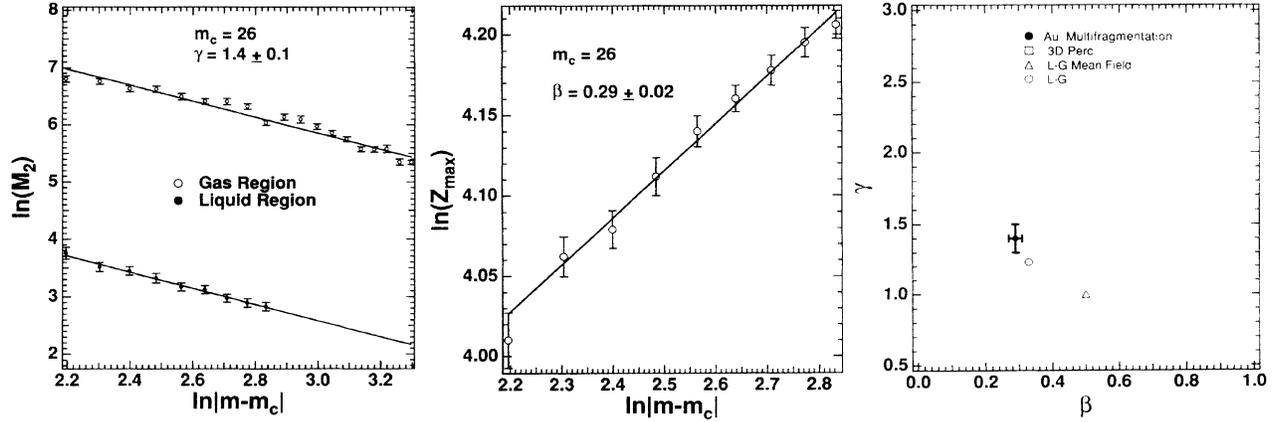

**FIGURE 23.** Left: An example of the determination of the critical exponent $\gamma$ for particular vapor and liquid fitting regions; the natural logarithm of the second moment of the cluster distribution plotted as a function of the natural logarithm of the distance from the critical point as measured by the total cluster multiplicity. Middle: An example of the determination of the critical exponent $\beta$ for a particular liquid fitting region; the natural logarithm of the charge of the largest cluster plotted as a function of the natural logarithm of the distance from the critical point as measured by the total cluster multiplicity. Right: The exponent $\gamma$ versus the exponent $\beta$ values determined from clusters produced in the reaction 1 AGeV Au+C and for $d = 3$ percolation, $d = 3$ Ising and mean field[89], see plot for legend.

**TABLE 2.** Values for the critical exponents. The exponent values given by the ratio of integer numbers are known exactly. The exponent vales not followed by citations are determined via the scaling relations given above. The exponent values for nuclear matter are the average for results from experiments which measured exclusive cluster yields [89, 114, 129, 144, 159, 177, 178, 183, 192, 193, 189, 212].

|  | $\beta$ | $\gamma$ | $\sigma$ | $\tau$ |
|---|---|---|---|---|
| $d = 2$ **Ising** | $\frac{1}{8}$ [6] | $\frac{7}{4}$ [6] | $\frac{8}{15}$ | $\frac{31}{15}$ |
| $d = 3$ **Ising** | $0.3265 \pm 0.0001$ [184] | $1.237 \pm 0.002$ [184] | $0.6395 \pm 0.0008$ | $2.209 \pm 0.006$ |
| **Nuclear matter** | $0.324 \pm 0.008$ | $1.25 \pm 0.07$ | $0.63 \pm 0.02$ | $2.18 \pm 0.02$ |
| $d = 2$ **percolation** | $\frac{5}{36}$ [43, 44, 47] | $\frac{43}{18}$ [43, 44, 47] | $\frac{36}{91}$ | $\frac{187}{91}$ |
| $d = 3$ **percolation** | $0.418 \pm 0.002$ | $1.793 \pm 0.003$ | $0.4522 \pm 0.0008$ [142] | $2.18906 \pm 0.00006$ [142] |
| **mean field** | $\frac{1}{2}$ | $1$ | $\frac{2}{3}$ | $\frac{7}{3}$ |

and binomial for the case of finite systems (see Eq. (83)), is inherent in not only Fisher's theory, but any physical cluster theory which assumes that a non-ideal vapor can be approximated by an ideal vapor of clusters with the formation of clusters exhausting the non-idealities. Thus, the stochasticity implied by reducibility is present in physical cluster models where all clusters are completely independent of each other. Figure 22 shows the reducibility feature observed in the cluster distributions arising from a variety of nuclear reactions [85, 95, 101, 106, 115, 127, 135, 149, 154, 155, 159, 164, 173] as well as from percolation calculations [159], Ising calculations [161, 180, 198] and classical molecular dynamics calculations [214].

Using reverse kinematics, the clusters produced in the reaction of 1 AGeV Au+C were studied [89]. The moments of cluster charge distributions $M_k(m)$ were analyzed in a similar fashion to the percolation cluster distributions discussed above (the total charged particle multiplicity $m$ was used as the "control parameter" in lieu of the more standard bond probability $q$, temperature $T$ or excitation energy $E^*$) [59, 67, 71, 72, 88, 123]. In this case the location of the critical point is given by $m_c$, the total charged particle multiplicity of clusters produced when the system reaches the critical point. Similarly, the distance from the critical point is given by $\varepsilon = m_c - m$. Fisher's theory (specifically the steps that yields equations (25) and (27)) leads to

$$M_2 \propto |\varepsilon|^{-\gamma} \text{ and } Z_{\max} \propto \varepsilon^{\beta}. \tag{88}$$

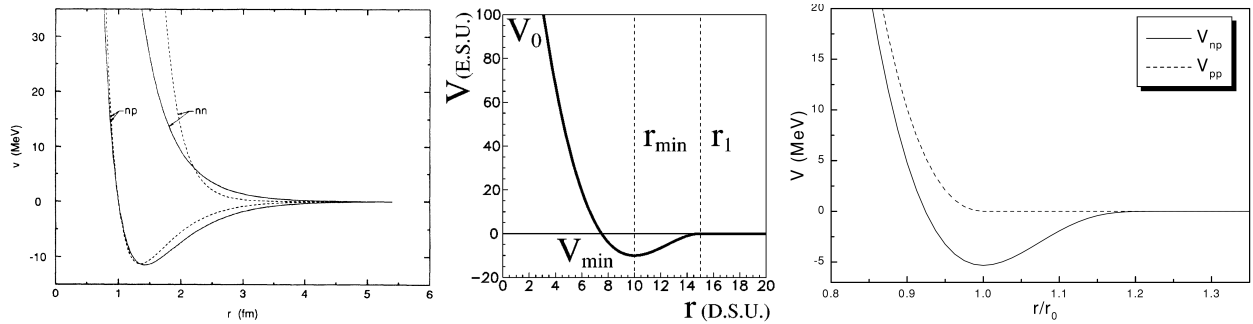

**FIGURE 24.** From left to right: an example of the Yukawa potential [77], an example of a Lennard-Jones type potential [182] and an example of a Lennard-Jones type potential with isospon [175]. All potentials share the short range repulsion (hard core) and longer range attraction.

With techniques developed and tested on percolation lattices [88, 123] the location of the critical point in terms of $m_c$ and the values of the critical exponents were measured from the exclusive cluster yields. Figure 23 shows the results of that analysis and Table 2 shows the critical exponent values from several different reactions. The similarity between the exponent values extracted from nuclear reactions and the values of the $d = 3$ Ising universality class is striking.

A variety of critiques of this analysis were discussed [102, 103, 104, 119] concerning the effects of mixing events of different temperatures by binning in multiplicity and the effects of including clusters produced in the collision in the analysis of clusters yields assumed to arise from an equilibrated source. Many of these criticisms were addressed in another analysis of this reaction with higher statistics which excluded clusters arising from the collision from consideration and studied the effects of binning percolation calculations in terms of cluster multiplicity [162]. In that work it was seem that the clusters produced in the initial collision had little effect on the extracted exponent values and that accurate critical exponents could be determined from lattices with as few as 216 sites when using cluster multiplicity as the control parameter. That work also stated the physical picture of cluster production from an equilibrated system [162]:

> Immediately following the collision, the gold projectile remnant is in an excited state with fewer than 197 nucleons. The excited remnant cools and expands and may evolve to the neighborhood of the critical point in the temperature-density plane, where clusters condense from a high temperature low density vapor of nucleons.

This physical picture and the analysis above raises several questions (beyond the fundamental question about how the system comes to equilibrium which has long been assumed to be the case [4, 5]). For instance, the simple power laws in Eq. (88) are valid so long as the chemical potential of the liquid is equal to the chemical potential of the vapor, i.e. the system is at coexistence: $\Delta \mu = 0$. Is there any evidence that the system is at coexistence? Where in pressure-temperature-density space is the system when the cluster's condence [130]? What is the meaning of density or pressure of a vapor which is not enclosed by any container? What are the effects of the nuclear nature of the system? Not only is there a cost in surface energy associated with the formation of a nuclear cluster (as shown in Fisher's theory), there is a cost in Coulomb energy, a cost in asymmetry energy, pair energy and so on. Whatever the answer to these questions, other types of analyses and various experiments measured similar exponent values [144, 177, 178, 183, 192, 193, 189, 212].

Another computational model that was used to study the phenomena of nuclear cluster formation was based on classical molecular dynamics attempted to study many of the questions above [61, 66, 69, 77, 90, 91, 99, 97, 109, 111, 118, 122, 124, 136, 141, 150, 175, 180, 182, 185, 186, 214].

Some calculations [61, 66, 69, 97, 111, 122, 124, 136, 141, 175, 180, 182, 185, 186, 214] were done either with a Lennard-Jones potential [3, 17, 22, 23] (modified or otherwise)

$$V(r) = 4E\left[\left(\frac{r_0}{r}\right)^{12} - \left(\frac{r_0}{r}\right)^6\right] \quad (89)$$

where $r$ is the distance between two particles and $E$ is the maximum depth of the potential well at $r = 2^{1/6} r_0$; for $r < r_0$ $V(r) \to +\infty$ ($r_0$ is the radius of the infinitely hard core) and for $r \gg r_0$ $V(r) \to 0^-$ (the long range attraction). Other calculations [77, 90, 91, 99, 109, 118, 150] were done using the Yukawa potential [11] (modified or otherwise)

$$\begin{aligned}
V_{nn}(r < r_{\text{cutoff}}) &= V_0 \left( \frac{e^{-\mu_0 r}}{r} - \frac{e^{-\mu_0 r_{\text{cutoff}}}}{r_{\text{cutoff}}} \right) \\
V_{np}(r < r_{\text{cutoff}}) &= V_r \left( \frac{e^{-\mu_r r}}{r} - \frac{e^{-\mu_r r_{\text{cutoff}}}}{r_{\text{cutoff}}} \right) - V_a \left( \frac{e^{-\mu_a r}}{r} - \frac{e^{-\mu_a r_{\text{cutoff}}}}{r_{\text{cutoff}}} \right) \\
V_{nn}(r > r_{\text{cutoff}}) &= V_{np}(r > r_{\text{cutoff}}) = 0
\end{aligned} \quad (90)$$

where $V_0$, $V_r$ and $V_r$ set the scale of the potentials and $\mu_0$, $\mu_r$ and $\mu_r$ give the range of the force. Figure 24 shows examples of the potentials used in various calculations. In general these efforts examined the clusters that were produced from systems with a few hundred particles enclosed in a container with periodic boundary conditions and/or a volume that was much larger than the volume taken up by the particles. Some calculations were performed by starting from an equilibrated drop of a few hundred constituents at a given temperature [61, 66, 91, 97, 99, 109, 111, 122, 124, 122, 141, 150, 175, 180, 185, 186, 214] while others started from two drops both near zero temperature, but which are excited through collisions [69, 77, 90, 118, 182].

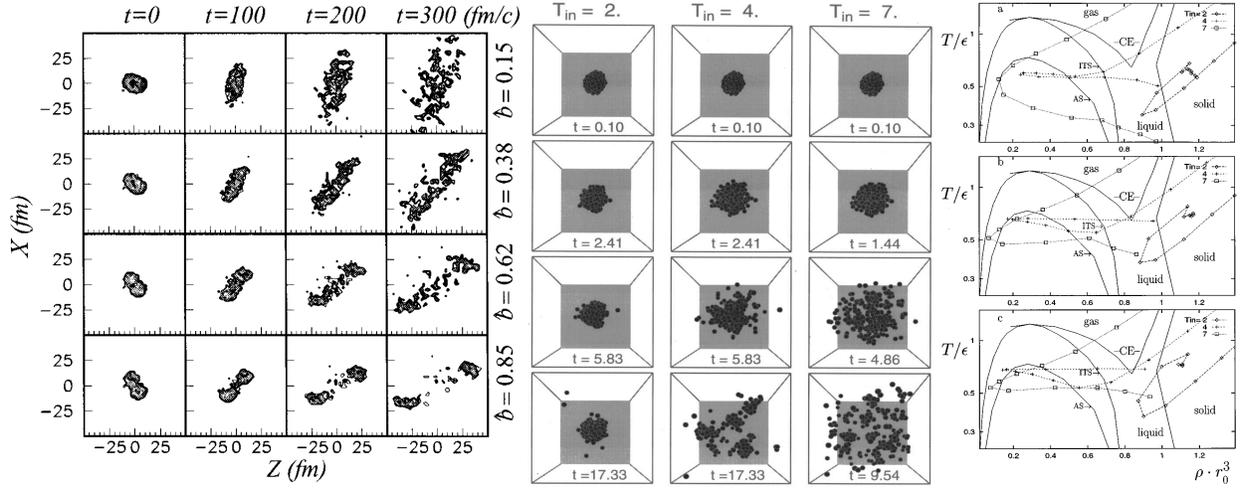

**FIGURE 25.** Left: density profiles of the dynamical evolution in coordinate space as a function of time and impact parameter for two colliding liquid drops [118]. Middle the time evolution of an $Ar_{300}$ cluster for various initial temperatures $T_{in}$: $T_{in} = 2.0$ shows a typical evaporation process; $T_{in} = 4$ shows the production of clusters of all sizes; and $T_{in} = 7.0$ shows a rapid expansion which may lead to instabilities and the formation of vapor clusters [124]. Right: the trajectory of the system in temperature-density space: from top to bottom the evolution of $Ar_{100}$, $Ar_{300}$ and $Ar_{500}$; solid curves indicate coexistence (CE), the isothermal spinodal (ITS) and the adiabatic spinodal (AS) for an infinite system. For all systems the temperature decreases with increasing time. The density shown for the trajectories is the central density of the largest cluster [124].

In general it was found the classical molecular dymanics calculations could reproduce, in quality, several features associated with experimentally measured clusters such as: the liquid-drop behavior of the binding energy [66, 182]; cluster yields (e.g. those shown in figures 13 and 14) which were also well described by Fisher's theory and Eq. (18) [69, 90, 91, 97, 99, 109, 111, 118, 122, 124, 141, 150, 185, 186, 214]; the Campi plots (shown in Fig. 18) [99, 118, 124, 175]; peaks in the moments of the cluster distributions and the associated critical exponents [186]; reducibility [180, 214]; and thermal scaling and the associated barrier dependence on cluster size [180]. Figure 26 shows some of these results.

While the features of the cluster distributions exhibited thermal and seemingly critical features, estimates of the trajectories (temperature and density as functions of time $T(t)$ and $\rho(t)$) of the systems studied rarely passed close to the liquid-vapor critical point [97]. For example, see the trajectories shown in Fig. 25 which shows that none of the trajectories considered pass near the liquid vapor critical point (while all trajectories pass near the adiabatic critical point) yet for $T_{in} \simeq 4$ critical behavior is reported [124].

One possible solution to this paradox is that the critical point of a system depends on the size of the system [24, 36, 37, 79, 94, 100]

$$T_c(\infty) - T_c(L) \propto L^{-1/\nu} \quad \text{and} \quad \rho_c(\infty) - \rho_c(L) \propto L^{-(d-1/\nu)}. \quad (91)$$

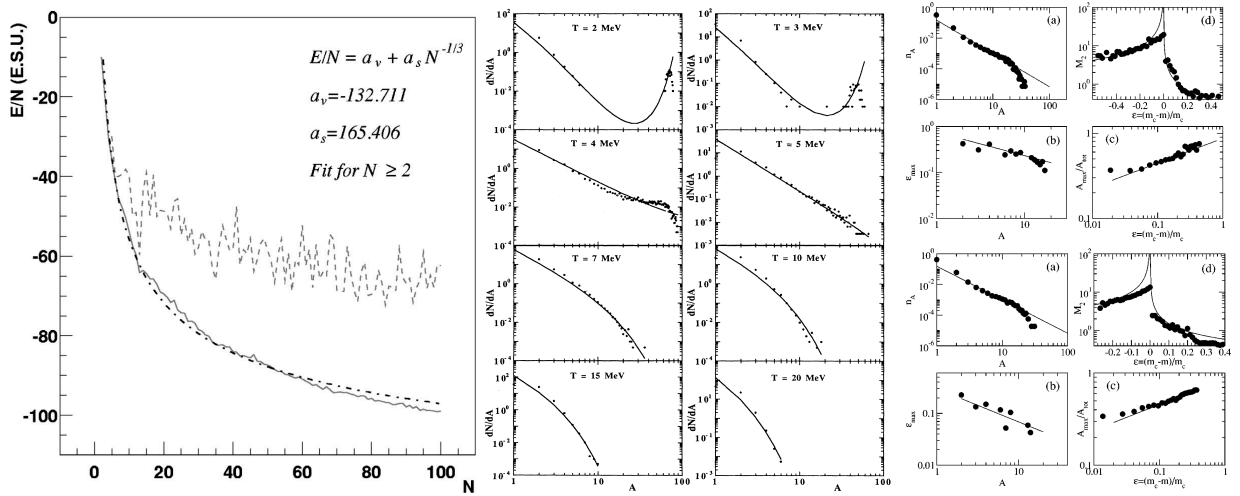

**FIGURE 26.** Left: The solid line shows the (binding) energy per particle of drops from a classical molecular dynamics calculation at low temperatures; the dash-dot line shows a fit with to the equation (and fit parameters) shown in the figure; and the dashed line shows the least bound particle in the drop [182]. Middle: cluster mass distributions for classical molecular dynamics calculations of 100 particles; dots show the results of the calculations and solid lines show a fit to Eq. (74) [99]. Right: top: results from the primary cluster distributions for the (a) cluster mass yield at an input temperature which gives the best fit to a power law (the line shows the fit of a power law with the result $\tau = 2.18 \pm 0.03$), (b) location of the peak in $A$-sized cluster production (the line shows the fit of a power law with the result $\sigma = 0.51 \pm 0.15$), (c) size of the largest cluster (the line shows the fit of a power law with the result $\beta = 0.29 \pm 0.08$)and (d) peaking behavior of the second moment of the cluster distribution (the line shows the fit of a power law with the result $\gamma = 0.77 \pm 0.25$); bottom: the same as the top but for the asymptotic time cluster distribution (here $\tau = 2.18 \pm 0.03$, $\sigma = 0.64 \pm 0.18$, $\beta = 0.28 \pm 0.13$ and $\gamma = 0.72 \pm 0.33$). Both the primary and asymptotic cluster yields give the same critical exponent values which are (expect for $\gamma$) similar to the $d = 3$ Ising values shown in Table 2 [186].

However, the size referred to in the scaling equations above, $L$, is the size of the volume in which the fluid is enclosed [100] and not the number of particles inside the volume. Thus, one may not see any such finite size scaling of the critical point if even just a few hundred particles are enclosed in a sufficiently large volume, or they enclosed in a volume with periodic boundary conditions (which lessens the effects of finite size [24, 36, 37]) or if they are not enclosed in any volume at all.

Another possibility is that the temperatures and densities used to construct trajectories as shown in Fig. 25 are not the pertinent quantities. Generally, the temperatures and densities used to construct such trajectories are associated with the central region of the largest cluster [61, 66, 69, 77, 90, 91, 97, 99, 124]. For instance, at low temperatures cluster production should be predominantly a surface phenomena, thus the temperature and density at the core of the evaporating cluster are less important than the conditions at or near the surface. In any case, it is clear that the clusters produced in classical molecular dynamical calculations appear thermal in nature, however it is still an open question how the dynamics leads to this result.

The Ising model, another well studied system [6, 9, 10, 24, 29, 35, 36, 37, 45, 49, 79, 109, 113, 121, 126, 128, 166, 176, 184, 198, 193, 199, 200, 209, 211], has also been widely used in the study of nuclear cluster production [93, 96, 110, 121, 133, 137, 138, 139, 143, 147, 148, 153, 161, 163, 168, 175, 179, 180, 181, 187].

The Ising model is a simple model of magnetic systems with spins $s_i$ placed on the vertices of a lattice. The spins are simplified models of atoms in a magnetic material and can point either up or down and can change direction based on the temperature $T$ of the system and the spin-spin interaction strength $J$. When the majority of spins point in a single direction (either up or down) the system has some net magnetization. When there are an equal number of up spins and down spins the system has no net magnetization. In infinite systems there is a single temperature (the critical temperature $T_c$) at which the phase changes from the magnetic phase to the non-magnetic phase.

The Hamiltonian $H$ of the Ising model has two terms: the interaction between nearest neighbor spins in a fixed lattice and the interaction between the fixed spins and an external applied field $H_{ext}$:

$$H = -J \sum_{i,j=\{n.n.\}} s_i s_j - H_{ext} \sum_i s_i. \tag{92}$$

In the absence of an external magnetic field, the system exhibits a first-order phase transition for temperatures up to the critical point at which it exhibits a continuous phase transition. The critical temperature for the $d = 2$ Ising model has been analytically for the square lattice to be $T_c = 2.26915\ldots J/k_B$ [6]. The critical temperature for the $d = 3$ Ising model has not been determined analytically; however, high temperature expansion techniques have yielded a value of $T_c = 4.51152 \pm 0.00004\ J/k_B$ [176] for the simple cubic lattice.

The zero-field Ising model is isomorphous with the lattice gas model [9, 10]. The positive spins are mapped to unoccupied sites in a lattice gas and the negative spins are mapped to occupied sites. The phase transition is then analogous to a liquid-vapor phase transition.

Typically realizations of the lattice are calculated as follows: for each lattice configuration, a random initial configuration of spins and a temperature is selected. Thermalization is reached via some algorithm, e.g. the Swendsen-Wang cluster spin-flip algorithm [75] using the Hoshen-Kopelman algorithm [**?** ] for cluster identification. After the system was thermalized, "geometric" clusters, i.e. nearest neighbor like spins, are identified (also using the Hoshen-Kopelman algorithm) and then the Coniglio-Klein algorithm [45] is used to break the "geometric" clusters into "physical" clusters. Use of the Swendsen-Wang algorithm and Coniglio-Klein clusters insure that the clusters are as close to the "physical" clusters observed in fluids as possible and do not suffer from problems such as the percolating critical point reached away from the thermal critical point or the presence of the Kertész line [73]. However, there are many different methods used to generate lattice realizations and many alterations made to the Hamiltonian or cluster definition in attempts to capture different sorts of physics. For instance, isospin [133, 148, 175, 180] and the Coulomb force [137, 168].